\documentclass{aastex631}

\newcommand{\kep}{\mbox{\textit{Kepler}}}
\newcommand{\ktwo}{\mbox{\textit{K2}}}
\newcommand{\T}{\mbox{\textit{TESS}}}

\newcommand{\muHz}{\mbox{$\mu$Hz}}

\newcommand{\dsct}{\mbox{$\delta$~Scuti}}
\newcommand{\gdor}{\mbox{$\gamma$~Doradus}}

\defcitealias{brogaard_age_2012}{2012}

\shorttitle{Photometry of NGC~6791 and NGC~6819}
\shortauthors{Colman et al.}
\graphicspath{{./}{}}

\begin{document}

\title{The \textit{Kepler} IRIS Catalog: Image subtraction light curves for 9,150 stars in and around the open clusters NGC~6791 and NGC~6819}

\correspondingauthor{Isabel Colman}
\email{icolman@amnh.org}

\author{Isabel L. Colman}
\affiliation{Sydney Institute for Astronomy (SIfA), School of Physics, University of Sydney, NSW 2006, Australia}
\affiliation{Stellar Astrophysics Centre, Department of Physics and Astronomy, Aarhus University, 8000 Aarhus C, Denmark}
\affiliation{Department of Astrophysics, American Museum of Natural History, 200 Central Park West, Manhattan, NY, USA}

\author{Timothy R. Bedding}
\affiliation{Sydney Institute for Astronomy (SIfA), School of Physics, University of Sydney, NSW 2006, Australia}
\affiliation{Stellar Astrophysics Centre, Department of Physics and Astronomy, Aarhus University, 8000 Aarhus C, Denmark}

\author{Daniel Huber}
\affiliation{Institute for Astronomy, University of Hawai`i, 2680 Woodlawn Drive, Honolulu, HI 96822, USA}

\author{Hans Kjeldsen}
\affiliation{Stellar Astrophysics Centre, Department of Physics and Astronomy, Aarhus University, 8000 Aarhus C, Denmark}



\begin{abstract}

The four-year \textit{Kepler} mission collected long cadence images of the open clusters NGC~6791 and NGC~6819, known as ``superstamps.'' Each superstamp region is a 200-pixel square that captures thousands of cluster members, plus foreground and background stars, of which only the brightest were targeted for long or short cadence photometry during the \textit{Kepler} mission. Using image subtraction photometry, we have produced light curves for every object in the \textit{Kepler} Input Catalog that falls on the superstamps. The IRIS catalog includes light curves for 9,150 stars, and contains a wealth of new data: 8,427 of these stars were not targeted at all by \textit{Kepler}, and we have increased the number of available quarters of long cadence data for 382 stars. The catalog is available as a high-level science product on MAST, with both raw photometric data for each quarter and corrected light curves for all available quarters for each star. We also present an introduction to our implementation of image subtraction photometry and the open source IRIS pipeline, alongside an overview of the data products, systematics, and catalog statistics.

\end{abstract}



\section{Introduction}
\label{sec:intro}

During the nominal four-year \kep\ mission, the telescope observed 150,000 stars in 30 minute long cadence \citep{jenkins_initial_2010}, and 512 per quarter in 1 minute short cadence \citep{gilliland_initial_2010}. Compared to the several million stars in the field, this places a limit on the completeness of statistical studies, with dependencies on factors such as magnitude and binarity \citep{wolniewicz_stars_2021}. The mission downloaded 3 full frame images (FFIs) per quarter, accessible via \citet{montet_long-term_2017}. This sampling cadence is not short enough to recover many kinds of periodicity among stars that were not individually targeted for long or short cadence photometry. Areas near the Galactic plane and in stellar clusters were under-targeted because the higher levels of crowding made it harder to isolate individual stars \citep{batalha_selection_2010}. Each targeted star was first downloaded as a series of images, known as a target pixel file (TPF), and a light curve was produced using simple aperture photometry (SAP). There is little we can do to extend the number of targets in the Galactic plane, apart from serendipitous finds in the background of TPFs, but the open clusters NGC~6791 and 6819 were targeted by ``superstamps,'' 200x200 pixel images (delivered as 20x100 ``sub-stamps''), which covered the clusters at the \kep\ long cadence of 30 minutes. Only a fraction of stars within these superstamps were targeted in the usual way, leaving many more still to be analyzed. We use image subtraction photometry, a method suited to crowded fields, to produce light curves for all 9,150 \kep\ Input Catalog (KIC) targets that fell on the superstamps.

We obtained these light curves using a novel implementation of image subtraction photometry, which we term IRIS (Increased Resolution Image Subtraction). The IRIS pipeline, outlined in Section~\ref{sec:imsmethods}, is open source and available on GitHub at \url{https://github.com/astrobel/kepler_iris}. In this paper, we present the \kep\ IRIS Catalog\footnote{\url{https://archive.stsci.edu/hlsp/iris}}, a collection of \kep\ long cadence light curves for 5,342 stars in the field of NGC~6791 and 3,808 stars for NGC~6819. These include 8,427 KIC entries that were not targeted during the nominal \kep\ mission. All light curves are freely available as a high-level science product (HLSP) on MAST via \dataset[https://doi.org/10.17909/t9-w8fw-4r14]{https://doi.org/10.17909/t9-w8fw-4r14}.

With 13$'$20$''$ square coverage of each cluster across the superstamps, the majority of cluster members are covered by these data products. Open clusters are prime testing grounds for theories of stellar evolution, as all the stars in a cluster are the same age. The open clusters covered by the \kep\ superstamps are two of four open clusters within the \kep\ field of view (FOV), alongside NGC~6811 and 6866. All four clusters have been the focus of many studies throughout and beyond the lifetime of the \kep\ mission. In this section, we provide an overview of those studies that focused on NGC~6791 and 6819.

NGC~6791 is one of the oldest known open clusters, with an age of 8.2$\pm$0.3 Gyr \citep{mckeever_helium_2019}, and an unusually high metallicity of $+$0.313$\pm$0.005 dex \citep{villanova_ngc_2018}, indicating that it likely formed in the Galactic bulge. NGC~6819 has an age of 2.38$\pm$0.05 Gyr \citep{brewer_determining_2016}, making it also one of the older known open clusters. It has a metallicity of $+$0.10$\pm$0.04 dex \citep{carrera_open_2019}, close to the Solar value. Many general studies of NGC~6819 have focused on stellar rotation \citep{balona_kepler_2013}, particularly on the application of the period-age-mass relation \citep{meibom_kepler_2011, meibom_spin-down_2015}. Further studies of the cluster have shown the connection between lithium abundance and stellar rotation rates \citep{deliyannis_li_2019}. 

Red giants are a particular focus of \kep\ open cluster studies: with well-characterized oscillations, their seismic parameters can be used in the asteroseismic scaling relations to provide mass and radius estimates, and hence cluster age and distance estimates \citep{basu_sounding_2011}. Additionally, with an understanding of the red giant populations in each cluster, red giant seismic parameters have been used to confirm cluster memberships \citep{stello_asteroseismic_2011} and study red giant evolution \citep{corsaro_asteroseismology_2012}. Red giant seismology in the \kep\ open clusters has also been used to examine the role of metallicity in cluster evolution \citep{hekker_asteroseismic_2011} and to test the asteroseismic scaling relations \citep{stello_detection_2010, miglio_asteroseismology_2012}. In NGC~6791, \citet{miglio_asteroseismology_2012} used red giants to find that there is no extreme mass loss on the red giant branch, noting that NGC~6819 is too young to exhibit significant mass loss. Studies of the younger red giant population in NGC~6819 have used seismology to distinguish between red giant branch and red clump stars \citep{handberg_ngc_2017}, and focused on individual stars of interest, such as the Lithium-rich red giant KIC~4937011 \citep{anthony-twarog_lithium-rich_2013, carlberg_puzzling_2015}. Showing that there are still more members to be found, \citet{komucyeya_new_2018} identified a new red giant in NGC~6819, namely KIC~5112840.

\kep's light curves for NGC~6791 and NGC~6819 have yielded studies of many other stellar oscillators. NGC~6791 is old enough to have white dwarfs: \citet{kalirai_stellar_2007} showed that the white dwarfs in NGC~6791 did not evolve via the Helium flash mechanism, but rather as the result of mass loss on the extreme horizontal branch, which suggests that mass loss is more efficient in higher-metallicity environments. \citet{garcia-berro_white-dwarf_2011} went on to show that the white dwarf population of the cluster is largely the result of binary mergers. NGC~6791 is also host to a notable pulsating subdwarf B (sdB) star, KIC~2438324 \citep{pablo_exploring_2011}; a further two sdB stars were identified using \kep\ data and analyzed by \citet{reed_discovery_2012}. NGC~6819 contains several \gdor\ and \dsct\ stars \citep{talamantes_bright_2010, balona_kepler_2013}. A notable classical pulsator in NGC~6791 is the \gdor\ binary KIC~2438249, which \citet{li_effect_2020} analyzed using pre-release data from this study.

There has also been a wide range of research into binary star systems in the \kep\ open clusters. In particular, eclipsing binary (EB) modeling provides reliable masses and radii; this allows us to determine stellar age, and in turn cluster age. In NGC~6791, \citet{grundahl_new_2008} used ground-based data to study the detached eclipsing binary KIC~2437452, providing cluster age and distance measurements. \citet{brogaard_age_2011} introduced three new EB models in the cluster, which they used to calibrate age and He abundance measurements; of these stars, \citet{hoyman_analysis_2019} went on to model the EB KIC~2438061 using \kep\ data, proving a distance estimate to the cluster, though their mass estimates differ from \citet{brogaard_age_2011}. NGC~6819 has proved host to a greater number of binaries than expected, as shown in the preliminary studies for this data release \citep{colman_thesis_2020}, which we will discuss in greater detail in Section~\ref{sec:lccomp}. Before \kep, there have been studies of binaries in NGC~6819 from the WIYN Open Cluster Survey (WOCS) \citep{talamantes_bright_2010, milliman_wiyn_2014}. Using \kep\ time series data, there has been increased focus on individual binaries to provide age and distance estimates, including the EB KIC~5024447 \citep{sandquist_long-period_2013}, spectroscopic binaries KIC~5113146 and KIC~5111815 \citep{soydugan_kepler_2020}, and the triple systems KIC~5113053 \citep{jeffries_wocs_2013} and KIC~5023948 \citep{brewer_determining_2016}. Many binaries in both clusters have also been identified as blue stragglers \citep[e.g.][]{hole_wiyn_2009, brogaard_blue_2018}. In recent work, \citet{vaidya_blue_2020} identified 75 blue stragglers in NGC~6791 and 29 in NGC~6819, while \citet{gao_5D_2020} found 49 blue stragglers in NGC~6791.

As this broad range of results shows, the \kep\ mission was uniquely suited to collecting large volumes of time series data for members of the open clusters in the FOV. However, the crowding in these regions placed limits on the completeness of photometric coverage across the open clusters. With this data release, we provide light curves for all KIC targets in NGC~6791 and 6819, including both cluster members and non-members, opening up new avenues of time domain study for both cluster analysis and general usage, such as the search for oscillating stars and exoplanets in the \kep\ field.

\section{Methods}
\label{sec:methods}

\subsection{\kep\ photometry}
\label{sec:kepphot}

The basic method for CCD photometry is aperture photometry, which involves selecting an aperture of pixels surrounding each target and adding the photon counts in those pixels (potentially with weighting) to calculate flux. The method can also involve defining an annulus around the aperture to estimate the background flux, which can be subtracted from each frame prior to photometry to create a more accurate time series. The main benefits of simple aperture photometry (SAP) are that it is easy to implement and fast to process; most complications in its implementation arise from defining the aperture. The SAP time series provided by \kep\ used apertures informed by the point spread function (PSF) and target brightness. The \kep\ PSF changed across the telescope's FOV, and as such it was modeled for each target individually. This led to apertures of varying sizes, as described by \citet{bryson_kepler_2010}. There is also an official \kep\ data product of detrended time series \citep{stumpe_kepler_2012, smith_kepler_2012, thompson_kepler_2016}, using the subtraction of cotrending basis vectors (CBVs) to remove systematics and trends from spacecraft pointing and jitter. The CBVs were calculated to fit the most common systematic trends in the data, in particular long term trends that arise from pointing drift. This method is called pre-search data conditioning (PDC) because it occurs prior to the search for transits in \kep\ data. Its implementation follows after the background subtraction step of SAP: even with an accurate aperture, SAP retains systematics from the pointing of the telescope which collected the data.

SAP and PDCSAP are the default methods of time series data delivery for \kep\ and \ktwo\ targets, and produces high-quality data for a majority of stellar targets. Using the \kep\ target pixel files (TPFs), one can also perform custom photometry on any targeted star. This has been a particular imperative for \ktwo\ data, which are harder to detrend due to the six hour pointing drift. There are multiple sets of \ktwo\ time series data available from different pipelines \citep[e.g.][]{vanderburg_technique_2014, lund_k2p2_2015}, all hosted on MAST. There are also superstamp images available for open clusters in \ktwo\ \citep{cody_k2superstamp:_2018}, presenting further opportunities for custom photometry. Full-frame images (FFIs) and calibration data provide opportunities for innovative approaches to obtain time series data for bright targets \citep{pope_photometry_2016, white_beyond_2017, pope_k2_2019}, where SAP is less effective due to pixel bloom and smearing, and non-targeted stars.

As well as bright stars, SAP is less effective for faint targets, and crowded targets. In particular, when there is minimal separation between neighboring targets, or indeed multiple stellar signals contributing to one pixel's flux, a different approach is required. In the following sections, we will discuss image subtraction photometry as an alternative to SAP, and our implementation for performing custom image subtraction photometry on the \kep\ superstamps.

\subsection{Image subtraction photometry}
\label{sec:ims}

To extract light curves from the NGC~6791 and 6819 superstamps, we used image subtraction photometry. Throughout the literature, this method is also known as difference imaging, or differential image analysis; we refer to the method as image subtraction to distinguish from the false positive vetting procedure known as difference imaging \citep{bryson_identification_2013}. Image subtraction was introduced by \citet{tomaney_expanding_1996}, and refined into today's widely-used algorithm by \citet{alard_method_1998}. Image subtraction was proposed as a method to offset poor seeing in ground-based telescopes. \citeauthor{alard_method_1998} created a reference image from an average of the twenty frames with the best seeing, after realigning each frame in the time series so that the target is centered on the same point. Following this, they used a kernel convolution to match the point-spread functions (PSFs) between each frame in the time series and the reference image, and then subtracted the reference image from each frame. This is the key step in the process: subtracting a reference image, which acts as an average, removes all non-variable signal in the time series. Following subtraction, \citeauthor{alard_method_1998} performed aperture photometry on each image, reconstructing any variable signal into a time series. This makes image subtraction an ideal method for studying stellar variability and probing crowded fields.

Since the introduction of the image subtraction method for photometry, its primary use has been on ground-based data. However, various image subtraction algorithms have been developed for use on \kep\ and \T\ pixel data. The most common is the \verb:FITSH: package \citep{pal_fitsh-_2012}, which includes utilities that have been used by \citet{soares-furtado_image_2017} and \citet{wallace_search_2019} on \ktwo\ cluster data, and by \citet{bouma_cluster_2019} in the search for exoplanets in \T\ open clusters. There is also the DIA pipeline \citep{oelkers_precision_2018}, which is designed to perform image subtraction photometry on \T\ data. Another method that has been used for crowded-field photometry in \kep\ is the causal pixel modeling (CPM) method \citep{wang_pixel-level_2017, hattori_unpopular_2021}, an approach to aperture photometry that defines an image background based on random pixels selected in the star's vicinity.

In this study, we introduce a new method for image subtraction photometry on space-based time series image data. Our implementation differs from previous methods in that we do not perform the convolution step for image realignment. We create a reference frame by taking an arithmetic mean of all frames in the time series, following a process of realignment to ensure that the target is centered on the same point in all frames. Alignment is a key issue when dealing with images from spacecraft, since we need to account for pointing drift. Our implementation of the algorithm uses a process of resampling and realignment to correct for this. The resampling step uses bilinear interpolation to redistribute flux into a sub-pixel structure --- leading to the term ``increased resolution image subtraction'' (IRIS) --- and the realignment step involves a further interpolation, this time using a centroid measurement to redistribute the flux accordingly. We present a detailed outline of this method in the following section.

\subsection{The IRIS algorithm}
\label{sec:imsmethods}

We performed image subtraction photometry on one quarter of \kep\ superstamp data at a time. We will cover the specifics of the practical application of this algorithm on \kep\ superstamp data in Sections~\ref{sec:imstargets} and \ref{sec:imsrunning}. The steps are as follows:

\begin{figure}
    \centering
    \includegraphics[width=0.25\textwidth]{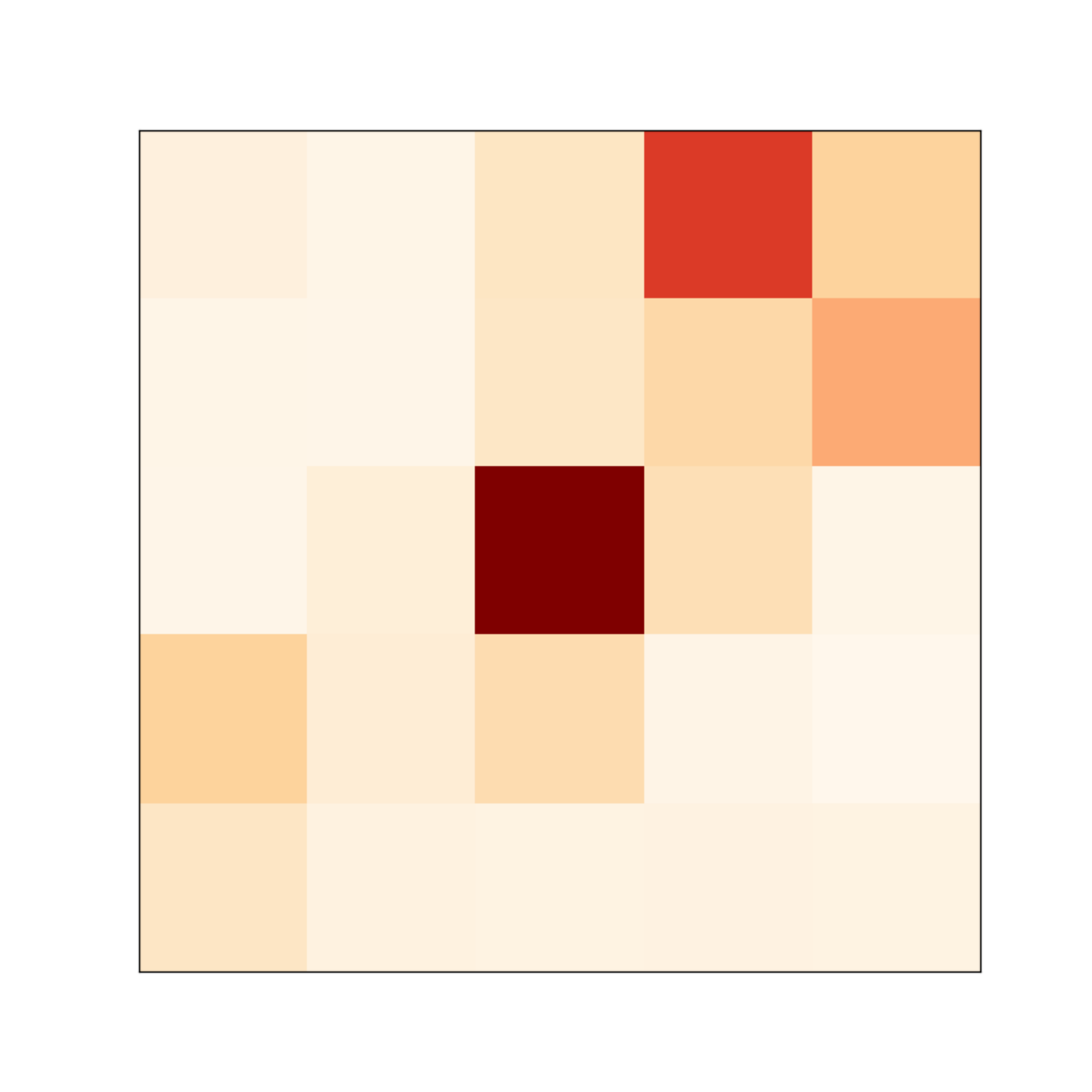}
    \caption{5x5 cutout example showing KIC~2437745, an ellipsoidal binary in NGC~6791.}
    \label{fig:cutout1}
\end{figure}

\begin{enumerate}
    \item Beginning with \kep\ superstamp data, we cut out a 5x5 pixel postage stamp centered on the target star, as in Figure~\ref{fig:cutout1}. This puts our data in a format akin to a \kep\ TPF --- that is, a time series of pixel images, here termed ``frames.'' We found the image centroid in pixel coordinates, calculated within a 3x3 pixel square around the target. We calculated the centroid using the equation for the first moment of an image,
    \begin{equation}
        \frac{\sum x_{i}f_{i}}{\sum x_{i}},
    \end{equation}
    where $x_{i}$ are the pixel coordinates along the $x$~axis and $f_{i}$ are the flux values in the corresponding pixels. We used a similar equation for the $y$~axis.
    
    In general, we found that the centroid shift for a \kep\ target is typically several hundredths of a pixel. Although small, this is significant, and consistent across all targets within the superstamps. We will discuss the centroiding process in greater detail in Section~\ref{sec:imscentroids}.
    \item We prepared the two-dimensional time series data for photometry by resampling each frame to have a large number of pixels per 4$''$ \kep\ pixel, as in the left panel of Figure~\ref{fig:cutout2}. Resampling allows for easier redistribution of flux according to the measured centroid. We included an interpolation at this stage to create a true distribution of fluxes across the new, finer resolution image. By inspecting the quality of light curves at various resampling factors, we found that a factor of 20 was sufficient for high quality photometric data.

    \begin{figure}
        \centering
        \includegraphics[width=0.25\textwidth]{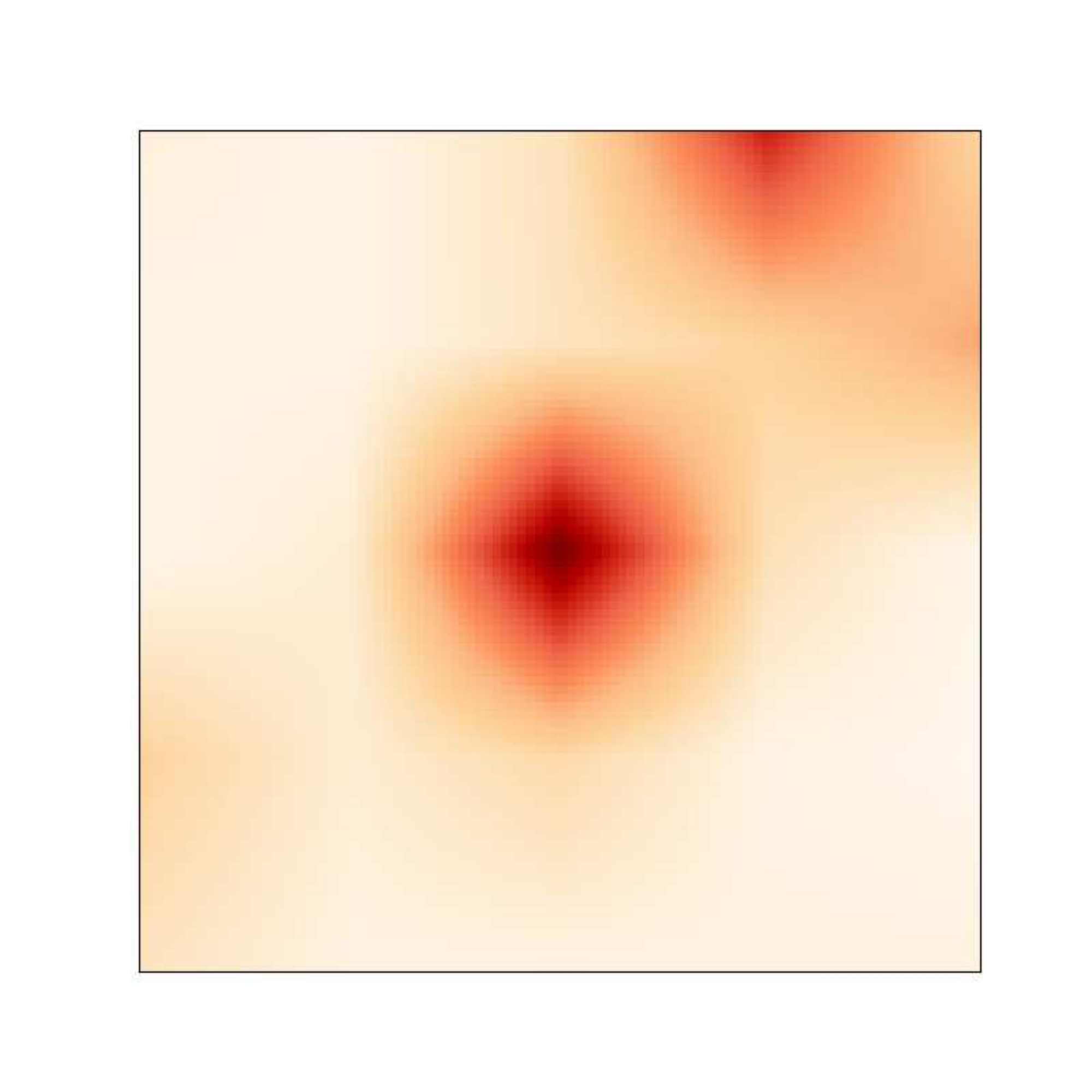}
        \includegraphics[width=0.25\textwidth]{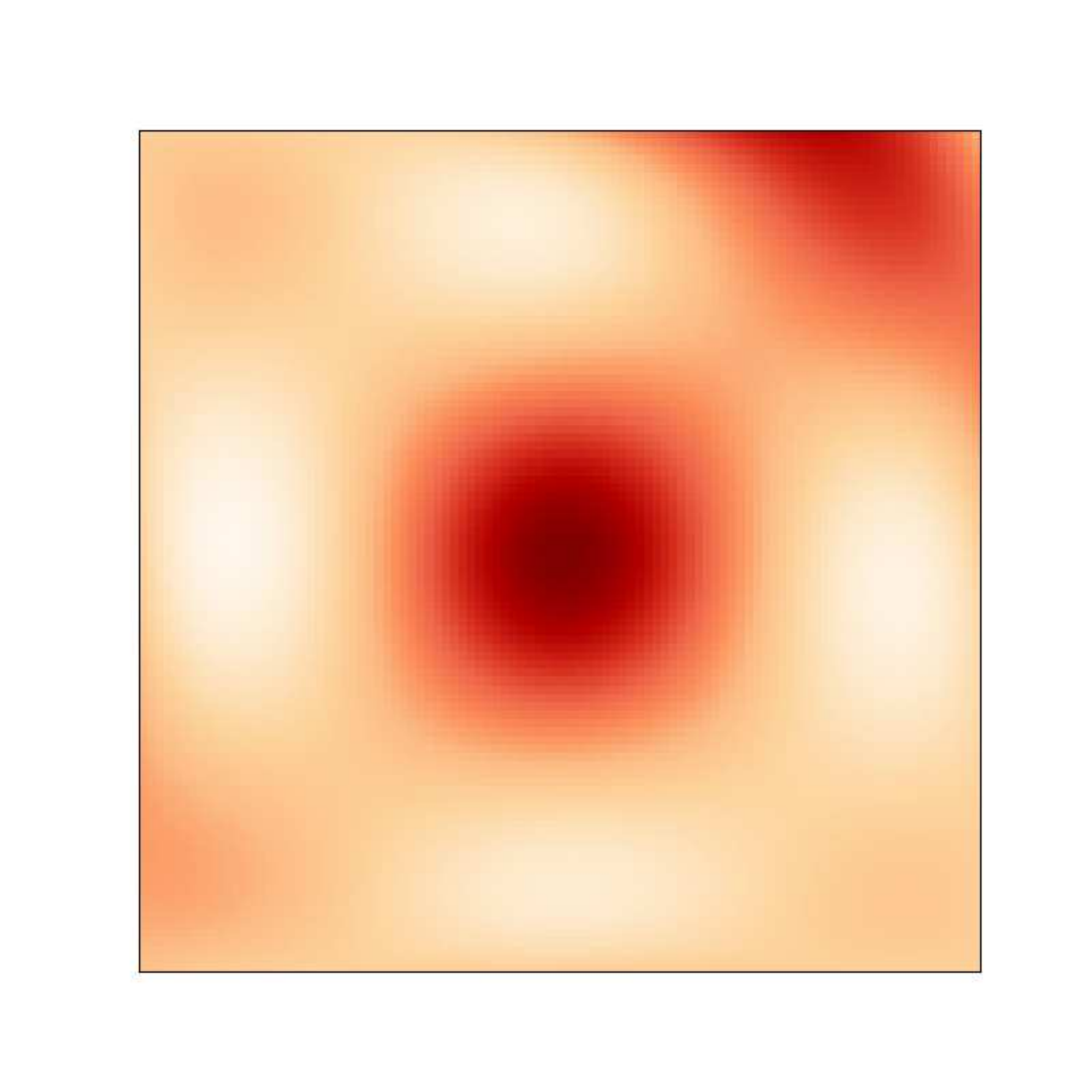}
        \caption{Left: 5x5 cutout of KIC~2437745, resampled using bilinear interpolation. Right: 5x5 cutout of KIC~2437745, resampled using bicubic interpolation. Note the haloing and edge effects, which led us to choose bilinear interpolation for this process.}
        \label{fig:cutout2}
    \end{figure}
    
    We performed the resampling using \verb:scipy:'s rectangular bivariate spline interpolation functionality \citep{virtanen_scipy_2019}. We used bilinear interpolation across both image axes. In general, cubic interpolation methods are preferred, as the cubic function is assumed to give a better fit. Bilinear interpolation is slightly less computationally intensive than bicubic interpolation; however, its main advantage in this study is that bicubic interpolation results in flux dilution due to a signal overshooting effect, meaning that bicubic interpolation leaves the interpolated image with a blurred look. Bilinear interpolation is better at preserving flux values, and retains a certain level of detail that bicubic does not, which is useful for understanding the structure of signal sources. Additionally, the \verb:scipy: implementation of bicubic interpolation results in pronounced edge effects at the borders of the 5x5 postage stamp. We show an example of this in the right panel of Figure~\ref{fig:cutout2}.
	\item Again using interpolation, we shifted each frame in the time series to move the measured centroid to the center of the postage stamp. We then took the average of all the realigned frames to use as our reference frame. We calculated the range of fluxes in the average frame, and selected the pixels with fluxes in the bottom 5\% of that range to determine a value for the background flux. We subtracted this background from each frame.
	\item Next, before performing the photometry, we defined an aperture around our target. Image subtraction isolates and measures variability, but in the case where multiple variable sources are nearby in a crowded field, it is necessary to limit the aperture to avoid the blending of multiple signals. The apertures we used are 2-D arrays with the dimensions of the average frame, with pixel values from 0 to 1, indicating weighting. We will cover aperture selection in more detail in Section~\ref{sec:imsmasks}.
	\item Finally, we subtracted the reference frame from each individual frame. This removed any non-variable flux that may have been in the aperture and accentuated any variable signal. We show the result of this subtraction for one frame in Figure~\ref{fig:cutout3}. Then, we added together the flux from each individual pixel, weighted by the aperture mask, making a single point in our time series data. The unit of these points is an arbitrary flux count, proportional to the square of the resampling factor and the original electrons per second count in each pixel prior to resampling and realignment. We repeated the subtraction and subsequent aperture photometry process for each frame. Finally, we added the aperture sum to each point in the time series, giving the time series point a flux value. Figure~\ref{fig:egfinalflux} shows the uncorrected time series for the example star used throughout this section.
    \begin{figure}
        \centering
        \includegraphics[width=0.35\textwidth]{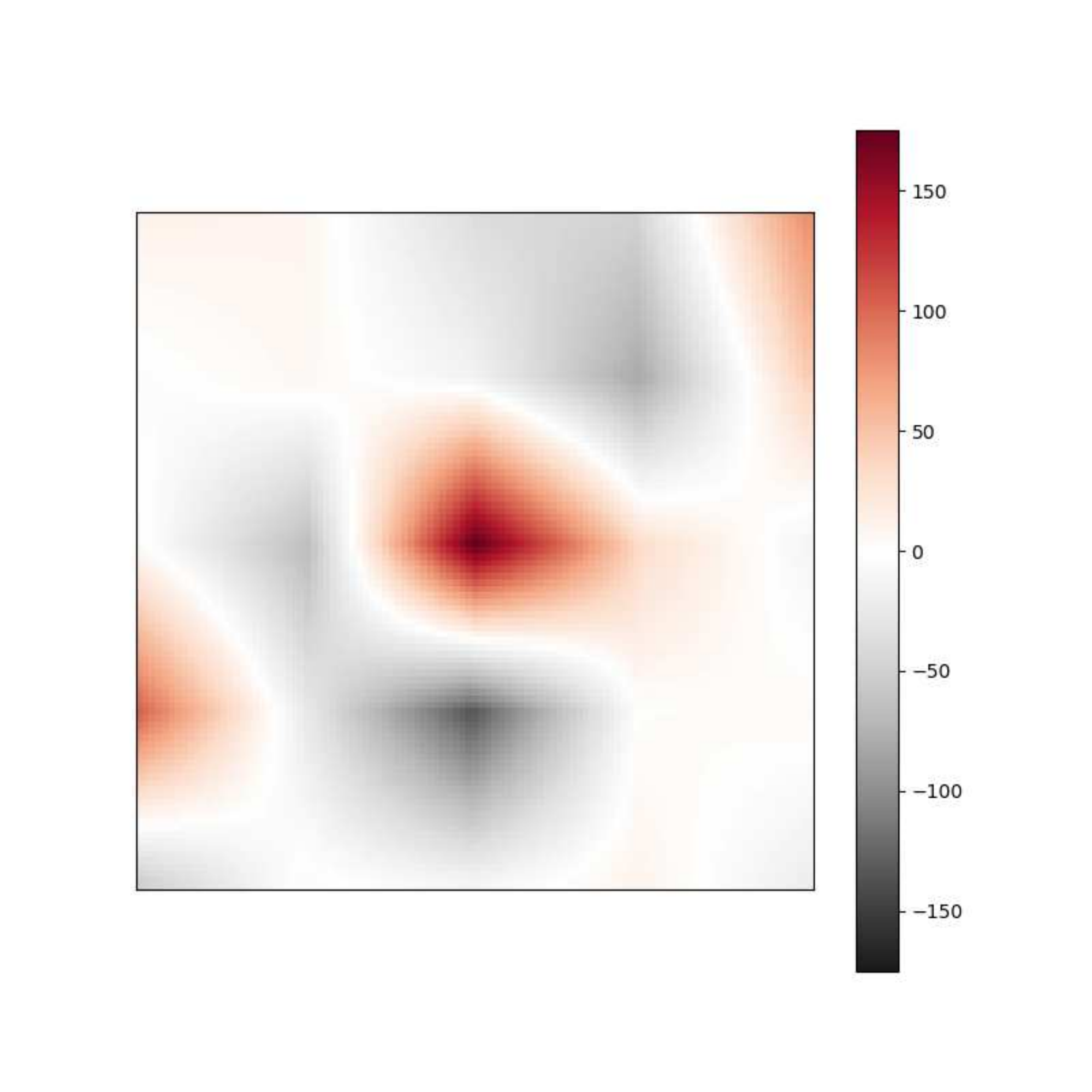}
        \caption{One frame of data for KIC~2437745, showing the results of image subtraction. Note that there is high amplitude variability from the target, as well as a companion star to the top right. Multiple variable stars in the same aperture can lead  to blending when using image subtraction photometry, so we select apertures to avoid this, as discussed in Section~\ref{sec:imsmasks}. The low amplitude positive and negative variation is due to small fluctuations between the individual frame and the average.}
        \label{fig:cutout3}
    \end{figure}
    \begin{figure*}
        \centering
        \includegraphics[width=\textwidth]{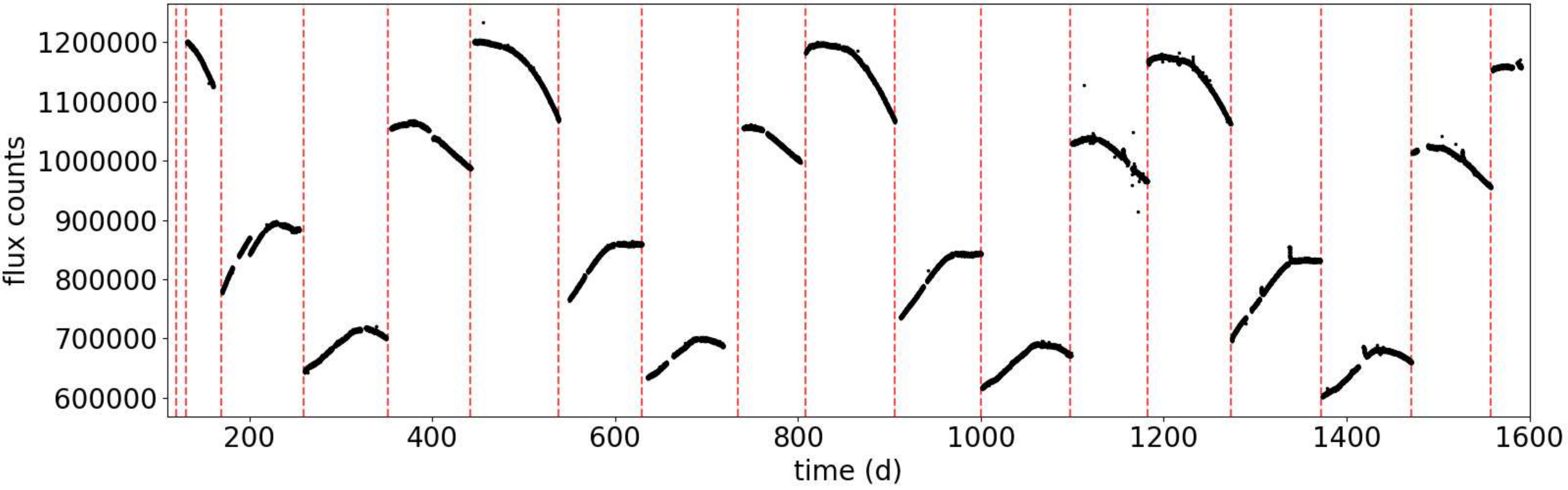}
        \caption{Uncorrected light curve of KIC~2437745, produced using image subtraction photometry. Dashed lines indicate quarter boundaries.}
        \label{fig:egfinalflux}
    \end{figure*}
\end{enumerate}

We implemented this algorithm in Python, using AstroPy for FITS file handling \citep{collaboration_astropy:_2013, collaboration_astropy_2018}, Pandas for database handling \citep{reback_pandas_2021}, NumPy for array handling \citep{harris_array_2020}, SciPy for interpolation \citep{virtanen_scipy_2020}, and Lightkurve for data management \citep{collaboration_lightkurve:_2021}. The code used to produce the IRIS light curves can be found at \url{https://github.com/astrobel/kepler_iris}.

\subsection{Cluster centroids}
\label{sec:imscentroids}

The image subtraction algorithm presented by \citet{alard_method_1998} uses a kernel convolution method to account for imperfect seeing from ground-based data. As \kep\ is a space telescope, this step is not necessary. However, we still need to ensure alignment between the reference frame and each individual frame. The benchmark image subtraction algorithm for \kep\ and \T\ data, \verb:FITSH: \citep{pal_fitsh-_2012}, retains the convolution step to perform this realignment, although it does not involve resampling. Alongside resampling, we introduce a centroiding process, which allows us to use interpolation to redistribute flux on a finer pixel grid. The overall change in centroid position at each cadence over one quarter can be as much as 20\% of a \kep\ pixel. For the process of realignment, we take image centroid measurements in 3x3 pixel squares centered on the target, and convert them from pixel coordinates into distances from the middle of the 5x5 postage stamp. The centroid displacement between any two cadences is typically several hundredths of a pixel along each axis; therefore, to precisely realign each frames without interpolation, we would need to resample by a factor of 100. However, the computational time involved in this would be untenable as part of a pipeline designed to run on thousands of stars. We found that interpolation on a grid resampled by a factor of 20 yields high-quality results within an acceptable computational time.

\begin{figure}
    \centering
    \includegraphics[width=\textwidth]{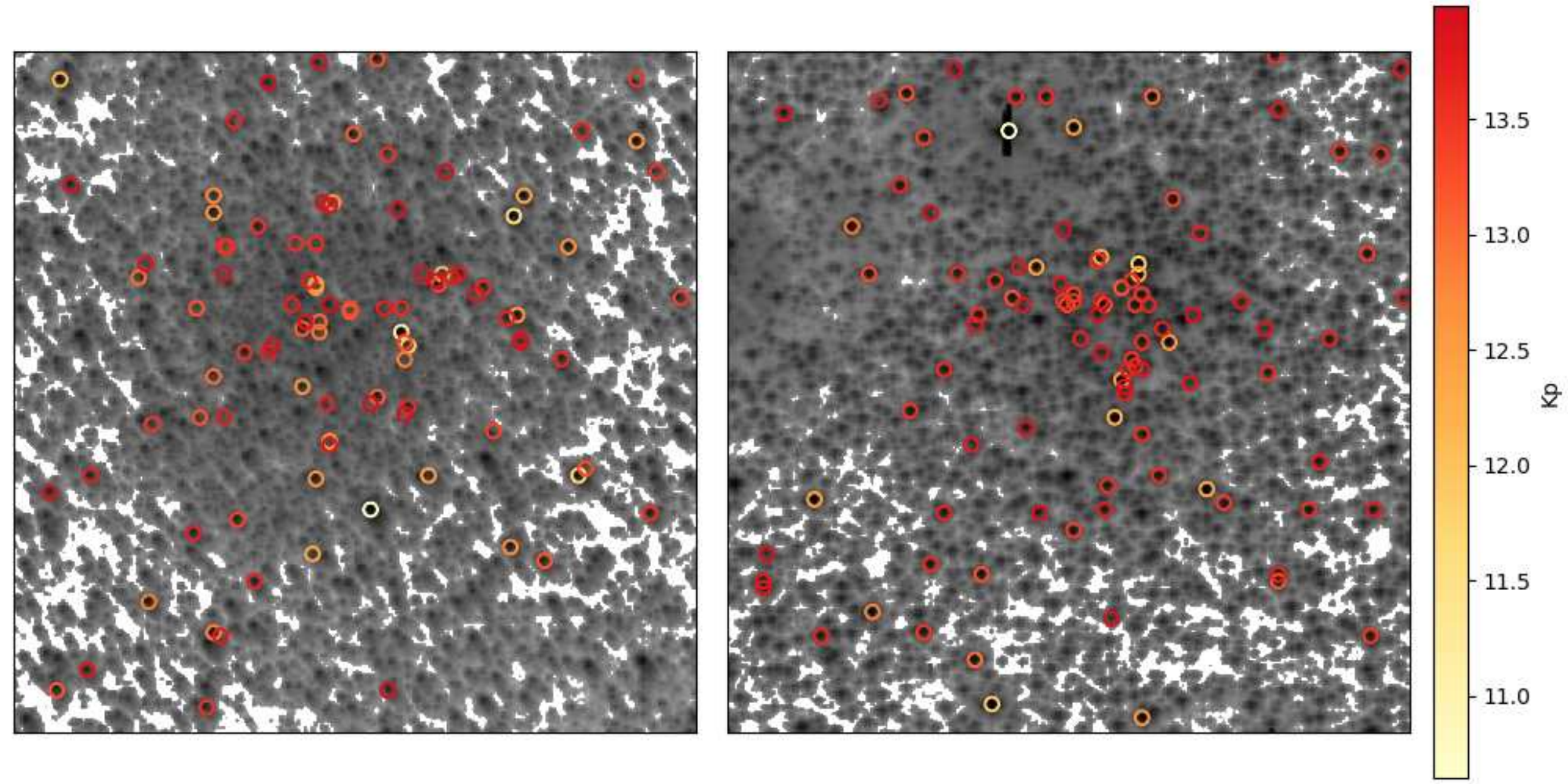}
    \caption{Stars used for ensemble centroid calculation in NGC~6791 (left) and NGC~6819 (right), plotted on images of the 200x200 superstamps. The color scale denotes \kep\ magnitude.}
    \label{fig:ensemble}
\end{figure}

In practice, we did not need to perform the centroid calculation for each target. Calculating the centroid of a star in a crowded FOV is unreliable because there are so many nearby sources of flux, many of which may be brighter than the target star and cause an incorrect measurement of the target centroid. To deal with these cases, we calculated an ensemble centroid for each cluster. We chose the 100 brightest targets in each cluster, discarding any stars which were missing quarters due to being near the edge of the superstamp, and stars where the 3x3 pixel centroiding region covered multiple 20x100 stamps. The distributions of these stars across the superstamps are shown in Figure~\ref{fig:ensemble}. We calculated the centroid measurements throughout each quarter for each of these stars, and created the ensemble centroid from an average at each cadence. Similar to the process of centroiding for individual targets, we then converted the ensemble centroid measurement into a pixel distance by subtracting the median. Note that centroid patterns repeat every four quarters, due to the orientation of the \kep\ telescope, meaning that the cluster is once again falling on the same CCDs. Note also that the patterns differ between the clusters: while systematics are similar enough that an ensemble centroid measurement is reliable within one 200x200 pixel region, the differences in detector systematics between the two clusters is sufficiently great that we could not use one ensemble centroid measurement for both clusters.

One advantage of the ensemble centroid is in computation time: the centroid calculation is the most time-intensive stage of the code, owing to the fact that it deals with individual pixels in tens of thousands of frames for each star. We reduced the computation time from $\sim$20~minutes per star with centroid calculation to $\sim$4~minutes per star with the pre-calculated ensemble centroid. As we will show in Section~\ref{sec:lccomp}, the ensemble centroid also proves to be better for faint and/or crowded targets, where a direct centroid measurement is easily skewed by nearby contaminants. Finally, we note that, since an ensemble centroid measurement is more computationally efficient, it will likely be highly useful for broader applications of the IRIS algorithm to large volumes of space-based data, such as \T\ photometry.

For the remainder of this paper, we refer to the use of the ensemble centroid measurements for the IRIS algorithm as EC-IRIS, and the use of individual target centroid measurements as TC-IRIS.

\subsection{Aperture selection}
\label{sec:imsmasks}

The apertures we used for this pipeline are weighted 2D Gaussian functions with the dimensions of the resampled 5x5 postage stamp. We adjusted the width of the Gaussian based on magnitude and crowding, which we represented with an isolation parameter. To obtain this parameter, we estimated magnitudes for each target using the unmasked and not-yet-resampled 5x5 postage stamp as our aperture, and then compared these to the catalog values of the \kep\ bandpass magnitude (Kp), calculated for every star in the \kep\ Input Catalog (KIC) using high-resolution ground-based photometry, and augmented with data from a range of all-sky surveys \citep{brown_kepler_2011}. To obtain the postage stamp magnitude ($M_{\rm 5x5}$), we summed the flux in each quarter and compared to a reference star, then took an average across all quarters. We calculated the flux for each quarter using

\begin{equation}
    M_{\rm 5x5} = M_{\rm ref} - 2.5{\rm log}_{10}\left(\frac{F}{F_{\rm ref}}\right).
    \label{eq:mags}
\end{equation}

\noindent We chose one isolated, bright target in each cluster to act as a reference star. The reference star for NGC~6791 was KIC~2436824, with Kp~=~14.515 and $F_{\rm ref}$~=~29292 counts, and for NGC~6819 we used KIC~5111949, with Kp~=~12.791 and $F_{\rm ref}$~=~106088 counts. Both stars are red giants.

We found that this process systematically underestimated the magnitudes for fainter stars --- that is, our calculated Kp was systematically brighter --- because the 5x5 aperture captured additional flux from neighboring stars. This information was used to measure a target's isolation in any given 5x5 postage stamp. Once we obtained $M_{\rm 5x5}$ for each target, we calculated an isolation measure $I$ using

\begin{equation}
    I = \frac{1}{2.5^{Kp - M_{\rm 5x5}}},
    \label{eq:iso}
\end{equation}

\noindent where 2.5 is an approximation for $10^{0.4}$, which is sufficient for the application of the isolation measure. By subtracting $M_{\rm 5x5}$ from the catalog Kp, converting to a brightness ratio and then inverting this value, we ensure that $I>1$ when $M_{\rm 5x5} >$~Kp, and $0<I<1$ when $M_{\rm 5x5} <$~Kp. In other words, $I<1$ indicates that we have overestimated the target's brightness, and therefore that we must use a smaller aperture to reduce the effects of crowding.

\begin{figure*}
    \centering
    \includegraphics[width=\textwidth]{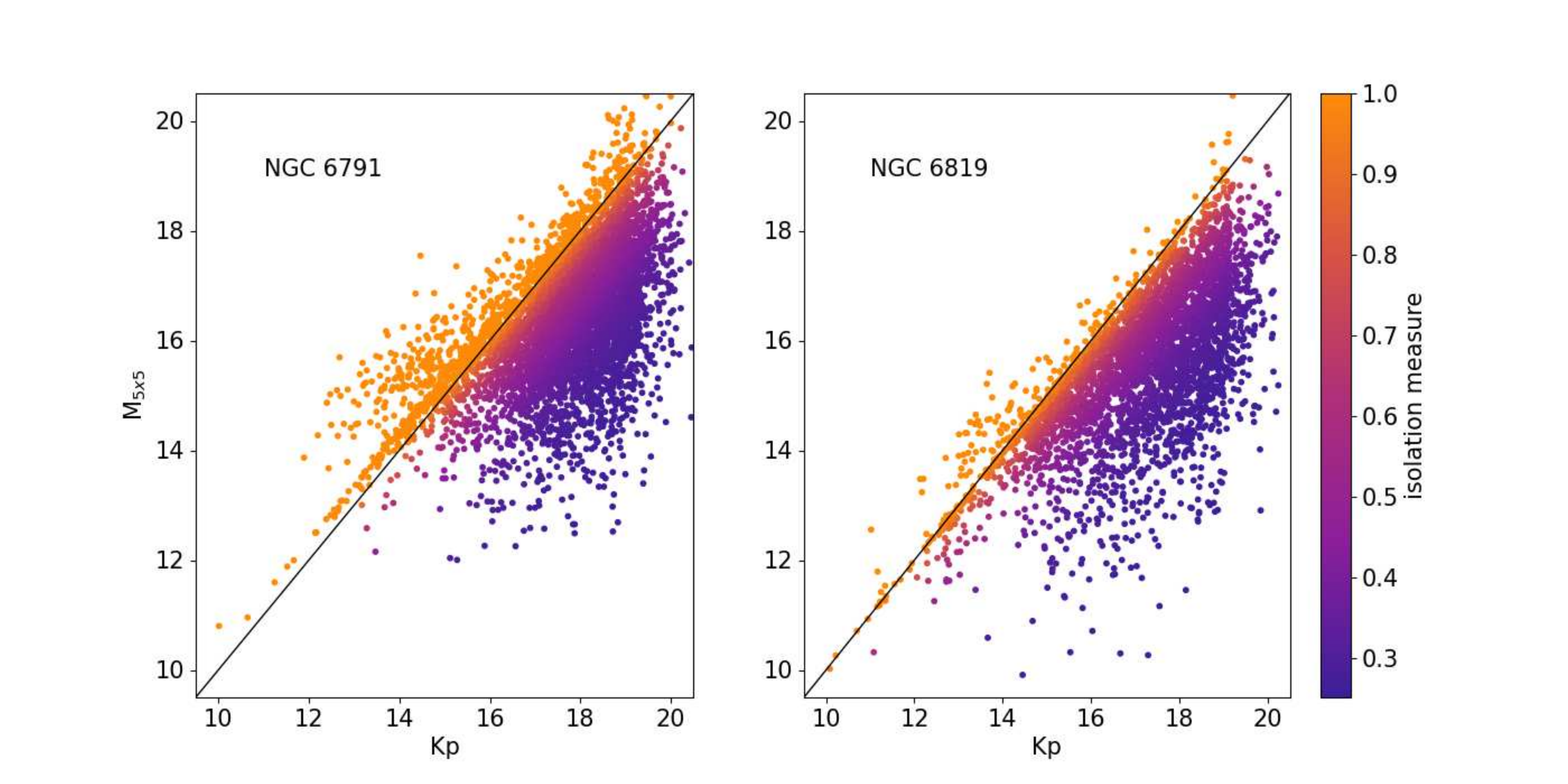}
    \caption{The magnitudes calculated in this study compared to magnitudes from the KIC, with a one-to-one line shown in black. Points are color-coded by the adjusted isolation measure, which we used to scale weighted Gaussian apertures for photometry.}
    \label{fig:kpkp}
\end{figure*}

The basic aperture for photometry was a normalized 2D Gaussian weighted pixel aperture, with a variance (or $\sigma$) of 1 \kep\ pixel. We scaled the aperture by multiplying $\sigma$ by an adjusted isolation measure. For this process, we placed two constraints on $I$: first, we set all targets with $M_{\rm 5x5}>$~Kp to $I=1$. We found that too large an aperture introduced excess noise to the resulting light curves, often at the expense of signal. As such, we found that the standard aperture as described above was sufficient for all fairly isolated stars, no matter their magnitude. For targets with $M_{\rm 5x5}<$~Kp we implemented a second constraint, scaling $I$ to lie within the range $0.25<I<1$. This avoided producing too-small apertures for extremely crowded stars, such as those with $M_{\rm 5x5}~-$~Kp~$>2$. Figure~\ref{fig:kpkp} shows $M_{\rm 5x5}~-$ vs Kp for all targets across both clusters, color-coded by adjusted isolation measure.

Following the creation of an aperture, we multiplied each subtracted frame in the time series by the Gaussian aperture, which scaled the fluxes according to pixel weighting. For crowded targets with $M_{\rm 5x5}<$~Kp, the adjusted aperture decreased the weighting on pixels subject to contamination. Due to the logarithmic nature of magnitudes and hence the isolation measure, the base aperture did not change perceptibly for targets where $M_{\rm 5x5}$ is close to Kp. Only 810 targets in NGC~6791 ($\sim$15\%) and 166 targets in NGC~6819 ($\sim$4\%) were processed with no change to the base $\sigma=$~1 \kep\ pixel aperture; as is expected in open clusters, the remaining stars were subject to brightness contamination due to crowding, and processed with smaller apertures. The relatively larger percentage of $\sigma=$~1 \kep\ pixel apertures used for targets in NGC~6791 is due to lower levels of crowding in that cluster.

\subsection{Target selection}
\label{sec:imstargets}

We used the \kep\ Input Catalog (KIC) \citep{brown_kepler_2011} as our base catalog. From the MAST target search, we obtained a list of all targets in a 14$'$ radius of the central coordinates of each cluster, in order to perform image subtraction photometry on every possible target in the NGC~6791 and NGC~6819 superstamps. We only retained stars that lie within the superstamps (which are 200 \kep\ pixels square, or 13$'$20$''$.) Due to shifts between quarters of up to five pixels, targets towards the edges of the superstamps may only appear for a subset of all quarters.

This process resulted in 5,342 targets in the field of NGC~6791 and 3,808 targets for NGC~6819. There are 17 quarters of superstamp data for NGC~6791, and 14 for NGC~6819. Of the targets included in this data release, 91\% of NGC~6791 targets fall on the superstamps for all 17 quarters, and 93\% of NGC~6819 targets have 14 quarters of data. We also compared data availability in these regions to targets which have \kep\ long cadence data: 310 targets for NGC~6791 ($\sim$6\%) and 413 for NGC~6819 ($\sim$11\%) were targeted during the \kep\ mission. Of these, only 155 targets for NGC~6791 and 186 for NGC~6819 have the maximum available number of quarters; many also include Quarter 0, the commissioning quarter, for which there are no superstamp data. Many of these targeted stars are not cluster members. Our overall corpus of light curves covers 9,150 targets in the \kep\ IRIS data release, of which 8,427 were not targeted for long cadence photometry during the \kep\ mission. We also provide light curves with more of quarters for 382 \kep\ long cadence targets.

\subsection{Running the pipeline}
\label{sec:imsrunning}

We ran the pipeline on these stars using high performance computing facilities at the University of Sydney, using GNU Parallel \citep{tange_ole_2018_1146014} to decrease runtime. We processed each star with both the EC-IRIS and TC-IRIS pipelines, and one custom simple aperture photometry (SAP) pipeline. This SAP pipeline was designed to closely mimic standard \kep\ SAP, but included resampling (without realignment) for fair comparison between data in apertures of the same dimensions.

The only \kep\ quality flag we took into consideration was for manually excluded cadences, bit 9 \citep{thompson_kepler_2016}. The \verb:QUALITY: column in the \kep\ open cluster sub-stamps covers quality issues across all 20x100 pixels, which leads to a quality flag for almost every cadence; additionally, some stars lie on the corners of these 20x100 sub-stamps, creating further complications when we consider that quality flags ought to be drawn from all data used. We implemented a filter to remove only cadences with the ``manual exclusion'' flag, which indicates that the detector was uniformly affected by quality issues. Notably, this removed artefacts due to the coronal mass ejections in quarter 12 and 14 from our data.

\section{Results}
\label{sec:imsresults}

\subsection{Light curve processing and classification}
\label{sec:lcproc}

The first step in processing the light curves that result from the IRIS pipeline was cropping out safe modes and other few-cadence low-quality events, using a list of events supplied by the Sydney pipeline for \kep\ photometry \citep{huber_automated_2009}. As mentioned in Section~\ref{sec:imsrunning}) we removed the CME events prior to photometry; beyond these manual exclusions and the Sydney pipeline exclusions, some of the remaining cadences may be subject to quality issues, such as cosmic ray incidences and sudden pixel sensitivity dropout events.

\begin{figure*}
    \centering
    \includegraphics[width=\textwidth]{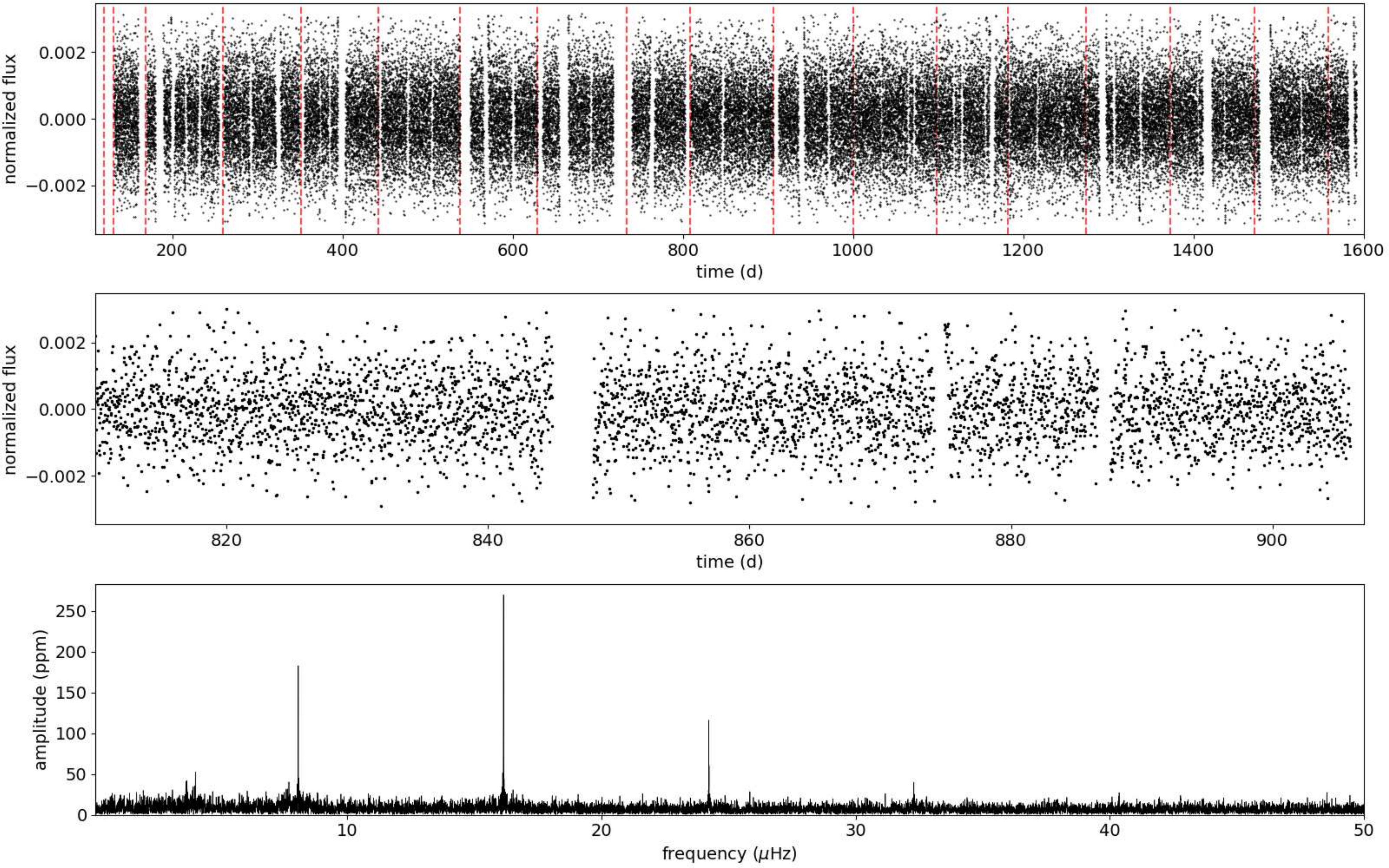}
    \caption{Top panel: corrected light curve for KIC~2437745. Middle panel: close-up of Quarter 9. Bottom panel: amplitude spectrum calculated using all 17 quarters of data, showing evidence of ellipsoidal variability.}
    \label{fig:2437745}
\end{figure*}

After the pipeline conducted photometry for each star, we performed light curve corrections to remove systematics. First, we applied a high-pass filter by smoothing with a Gaussian of width 100~cadences, followed by a 3$\sigma$ outlier clipping process. We performed the filter and clipping procedure in three iterations to each individual quarter. Normally, one would divide the light curve by the Gaussian curve used for filtering, and this would guarantee normalization. However, in crowded fields, the contribution of photon noise from contaminating stars is additive, meaning that a subtractive filter is more appropriate. The contaminating flux changes between quarters due to the differences in pixel sensitivity across different CCD modules. This led to significant variation in the scatter between time series points across different quarters, which introduced noise into the periodogram calculation. After subtracting a Gaussian curve from each quarter to account for both contamination and scatter, we calculated a mean flux. We then divided each quarter by the mean of those means, leading to more uniform light curves, such as the one shown in Figure~\ref{fig:2437745} for KIC~2437745, the example star from Section~\ref{sec:imsmethods}. This method worked well for the majority of stars; there are some targets with persistent systematics (see Section~\ref{sec:lcsys}) where custom corrections may need to be performed before further analysis.

For each target, we examined the image subtraction light curves for evidence of periodicity by computing a Lomb-Scargle periodogram \citep{lomb_least_1976}, as implemented by Astropy \citep{collaboration_astropy:_2013, collaboration_astropy_2018}. For purposes of categorization, we classified stars as either variable or non-variable based on the presence of significant peaks in the amplitude spectrum. One issue is the possibility of multiple variable signals, or the signal of a nearby bright contaminant, within one aperture. To manually check for contamination, we produced a periodogram of the light curves in each individual pixel of a star's 5x5 postage stamp \citep[see][]{bryson_identification_2013, colman_evidence_2017}. The majority of stars in our sample are faint enough that contamination must be present if the overall photometric signal is stronger in any pixel other than the central pixel of the 5x5 postage stamp. Due to the high completeness of this study, in most cases it was possible to identify the contaminant among the light curves of nearby targets in the dataset, confirming the pixel periodogram analysis method as highly reliable. There were only a small number of cases across the sample where two KIC targets occupied the same pixel and we detected signal that could come from one or both of the targets. In these and other ambiguous cases, we classified the target as variable.

\subsection{Systematics}
\label{sec:lcsys}

\begin{figure}
    \centering
    \includegraphics[width=0.7\textwidth]{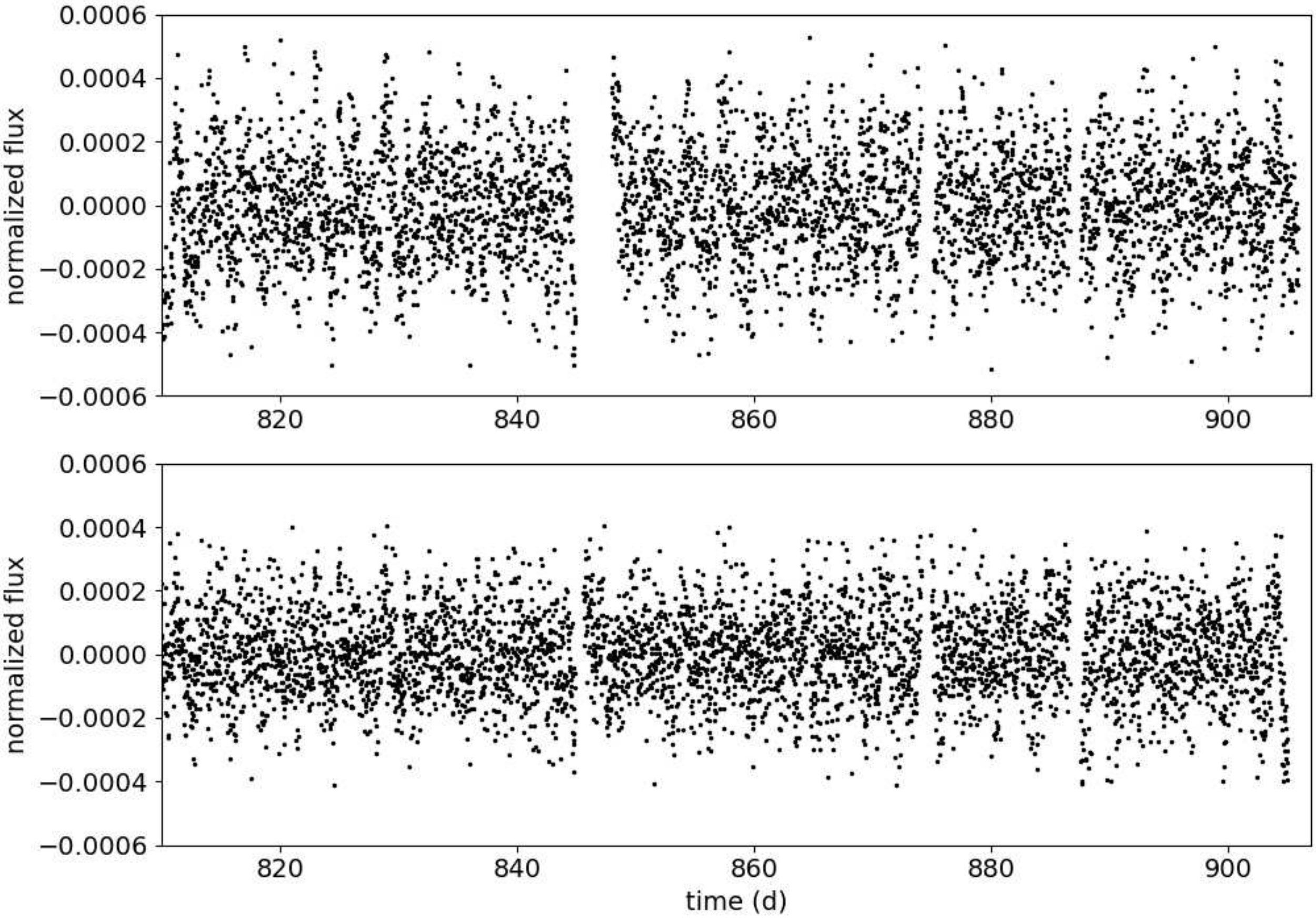}
    \caption{Quarter 9 EC-IRIS (top panel) and \kep\ SAP (bottom panel) data for KIC~5023743, a target which exhibits the spurious $\sim$3~d signal described in Section~\ref{sec:lcsys}. Note that the signal is present in both our light curve and a \kep\ SAP light curve, but appears to be higher amplitude in our data. This is due to the differing apertures; EC-IRIS apertures were optimized for faint targets.}
    \label{fig:spurious}
\end{figure}

A significant minority of these light curves display a $\sim$3~day signal which appears to be correlated with spacecraft motion. We have no firm explanation as to what causes this signal. There is no change in the signal with position on the detector, target brightness or isolation, or any intrinsic stellar properties. The presence of the signal is related to noise levels: the $\sim$3~d signal has a lower amplitude when it appears in targets which exhibit higher levels of shot noise. We have confirmed that this signal is also present in the \kep\ pipeline data for the same targets. Figure~\ref{fig:spurious} shows KIC~5023743, a star with \kep\ long cadence data available, where both light curves display the unknown $\sim$3~d signal. By percentage, the signal is more common among our light curves than in the \kep\ long cadence dataset; this is consistent with the higher density of field sampling in our dataset when compared with the sampling of stars targeted during the \kep\ mission.

Additionally, some targets in NGC~6791 show a spurious signal in Quarters~1~and~2 due to the influence of variable guide stars. For the majority of the \kep\ mission, the guide stars used for instrumental calibration were chosen to be non-variable; however, as documented in the \kep\ Data Characteristics Handbook \citep{van_cleve_kepler_2016}, there were two variable stars used for calibration in Quarters 0--2. One of these variable guide stars is an intrinsic variable with a period of 2.9~d, or $\sim$3.99~\muHz, and lies close to the NGC~6791 superstamps. This signal shows up in a small fraction of our NGC~6791 light curves. For studies of these data involving short period targets, we recommend the omission of Quarters~1~and~2. We also note that, despite the similar periodicity, this is unrelated to the unknown $\sim$3~d systematic described above.

\begin{figure}
    \centering
    \includegraphics[width=0.7\textwidth]{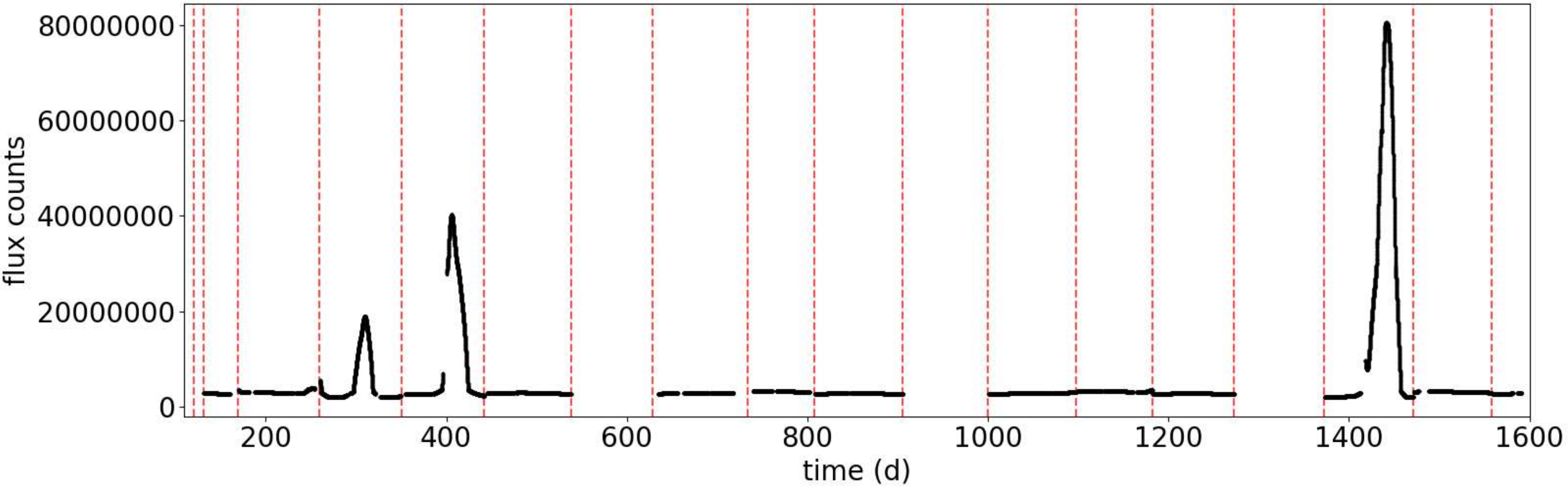}
    \includegraphics[width=0.7\textwidth]{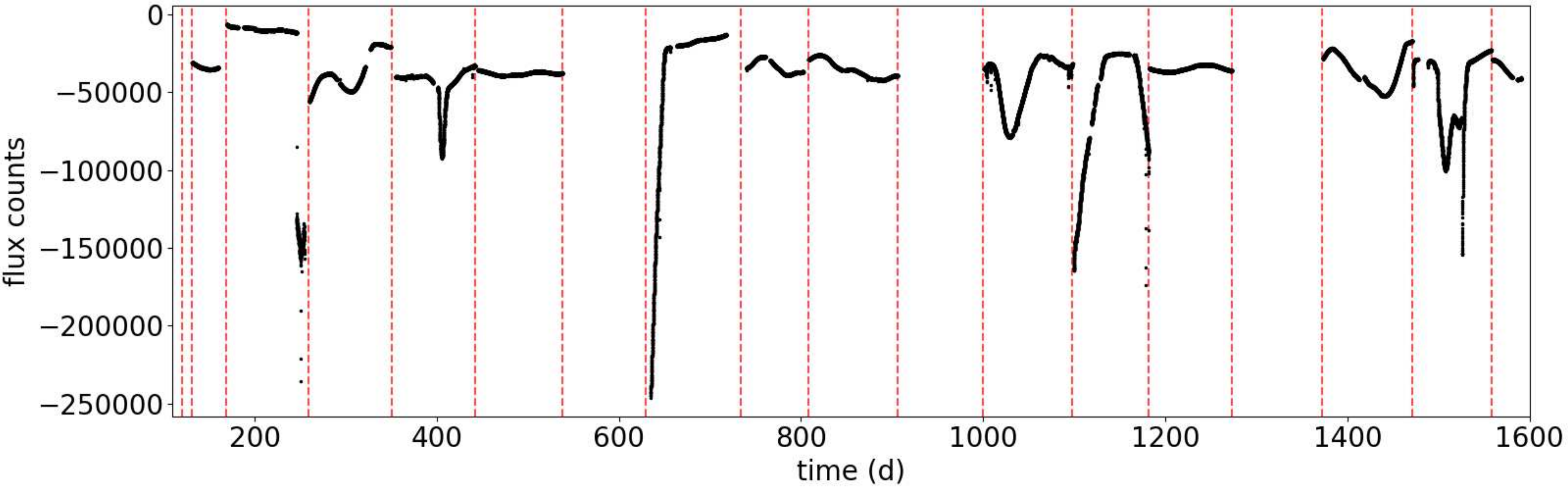}
    \caption{Top: KIC~5024943. Bottom: KIC~5025005. Both stars exhibit uneven variations in flux due to their proximity to a bright target. Dashed lines represent quarter boundaries.}
    \label{fig:notflares}
\end{figure}

Eight targets in NGC~6819 exhibit steep, irregular trends in their uncorrected light curves, with no uniformity across quarters. These variations occur across several orders of magnitude, and appear as either flare-like in shape, or as dramatic dips in brightness. We show two examples in Figure~\ref{fig:notflares}. In each case, the target lies in the close vicinity of one of two bright targets: KIC~5024470 (Kp~$=$~10.083) and KIC~5025003 (Kp~$=$~8.367). The targets and their contaminating star are listed in Table~\ref{tab:notflares}. As this is a pixel-level issue, we are unable to correct for this as part of the IRIS pipeline and basic light curve corrections. Nevertheless, the light curves are included in our data release for users to perform their own corrections.

\begin{table}
\begin{tabular}{rllll}
\multicolumn{1}{c}{\textbf{KIC}} & \textbf{RA} & \textbf{Dec} & \textbf{Kp} & \textbf{Contaminant}     \\
\hline\hline
5024943 & 295.38915   & 40.1768      & 14.854      & 5024470\\
5024964 & 295.392906  & 40.164344    & 16.716      & \\
5024979 & 295.395495  & 40.16895     & 17.291      & \\
5025005 & 295.3999995 & 40.167944    & 17.55       & 5025003 \\
5025024 & 295.40397   & 40.156006    & 16.663      & \\
5025051 & 295.408611  & 40.146194    & 18.079      & \\
5025063 & 295.4099295 & 40.15199     & 17.375      & \\
5025073 & 295.412547  & 40.146044    & 17.67       & \\
\hline
\end{tabular}
\caption{Targets in the field of NGC~6819 with sharp trends in their light curves due to uneven contamination by a nearby bright target.}
\label{tab:notflares}
\end{table}

\subsection{Pipeline statistics and method comparison}
\label{sec:lccomp}

Due to the limitations of the light curve corrections used and the unfeasibility of a one-size-fits-all corrections approach for this large dataset, we probably underestimate the number of low-frequency (or long-period) variable targets. With that in mind, we identified variability in 239 of 5,342 targets in the field of NGC~6791, and in 254 of 3,808 targets in the field of NGC~6819. The preponderance of variability in the field of NGC~6819 is noteworthy, as is the fact that this statistic is comparable to the 245 variables found in \citet{colman_thesis_2020}, where we studied a subset of these stars, 1,286 targets determined to be members of NGC~6819 \citep{drury_thesis_2020}. This suggests that NGC~6819 is host to an unusually large number of variable stars, or conversely that NGC~6791 may have a low number of variable stars.

\begin{figure*}
    \centering
    \includegraphics[width=\textwidth]{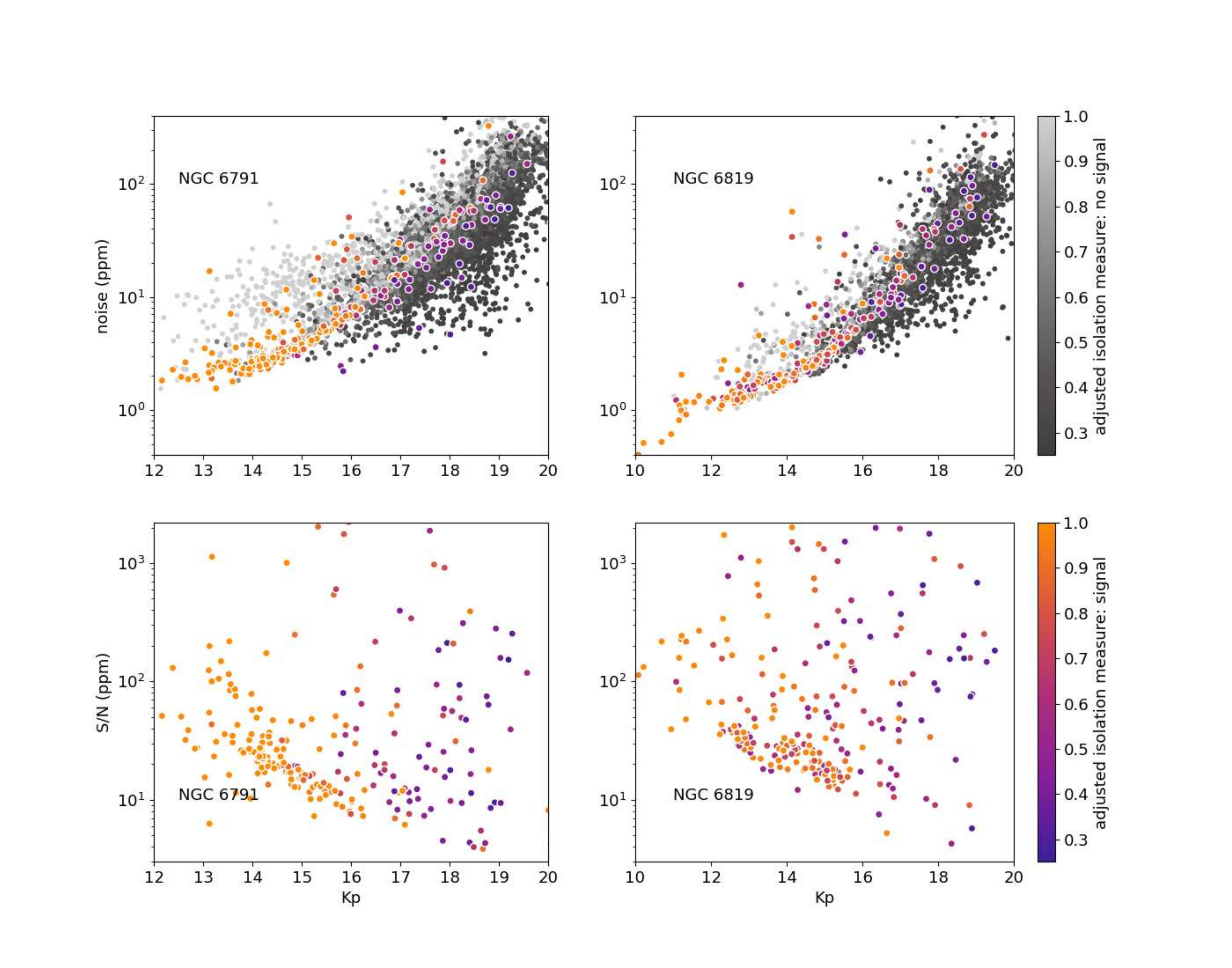}
    \caption{Top panels: noise vs Kp plots for both clusters. Bottom panels: signal-to-noise (S/N) vs Kp. NGC~6791 is on the left and NGC~6819 is on the right. All plots use noise and S/N measures as outlined in Section~\ref{sec:lccomp}, color-coded by adjusted isolation measure. In order to show a clear overview of trends, we set axis limits to exclude outliers in noise, S/N, and Kp. Note that the typical Kp ranges differ for each cluster. Note also that we do not plot the S/N ratio for non-variable stars, as it is determined based on the highest-amplitude peak of variability, which becomes arbitrary in an amplitude spectrum dominated by noise.}
    \label{fig:classifiedcmds}
\end{figure*}

Using the periodograms for each target, we measured the median of the amplitude spectrum between 250--280\muHz\ as a representative noise measure for each star, where $\sim283$\muHz\ is the Nyquist frequency for long cadence \kep\ data. We expected to find relatively few cases of variability above 250\muHz\ in this dataset, and comparatively many solar-like oscillators, which likely display a low-frequency red noise profile due to granulation; because of this we consider the higher frequencies a fair estimate of photon noise. We then used the noise measure to calculate a signal-to-noise (S/N) ratio, where the signal amplitude was taken as the highest amplitude present in the spectrum. Note that, despite the aperture sizes scaled to minimize excessive contamination by nearby targets, this method includes light curves subject to contamination, as described in Section~\ref{sec:lcproc}. Additionally, we see some crowded targets with artificially low noise due to high shot noise introduced at the final step of the image subtraction algorithm, the addition of the average flux. However, we have also managed to maximize the S/N ratios for these targets by our careful aperture choice, so we see almost no dilution in signal amplitudes below a strict lower limit that is a function of magnitude. Figure~\ref{fig:classifiedcmds} shows noise and S/N vs magnitude for both clusters.

\begin{figure*}
    \centering
    \includegraphics[width=\textwidth]{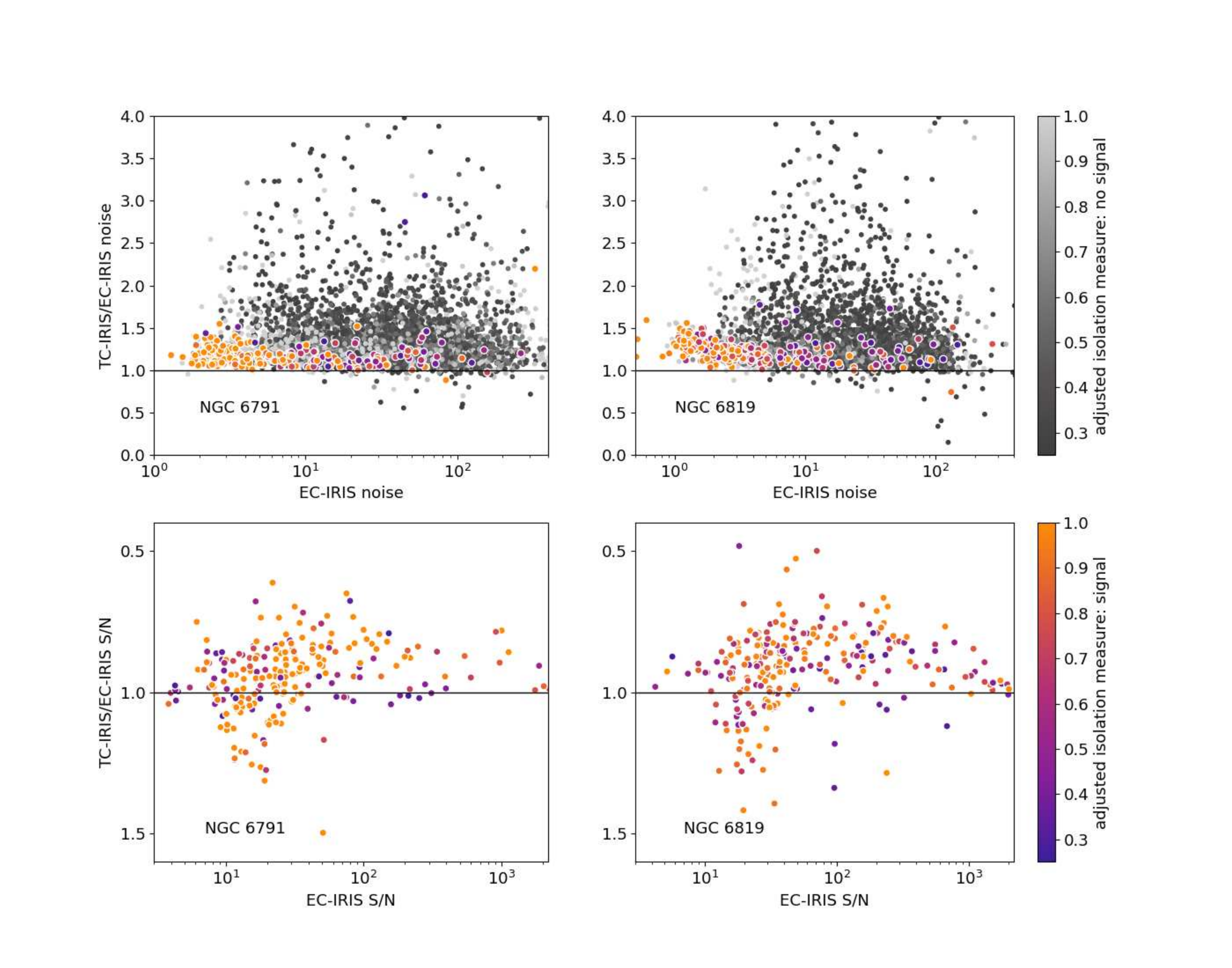}
    \caption{Left panels: NGC~6791. Right panels: NGC~6819. Top row: noise measures for the EC-IRIS and TC-IRIS pipelines, with EC-IRIS on the $x$~axis and the ratio between TC-IRIS and EC-IRIS on the $y$~axis. Bottom row: signal-to-noise measures plotted for the same two pipelines, with the y-axis inverted. For both noise and S/N, points will lie above the one-to-one line in the case of favorable performance by EC-IRIS, i.e. lower noise and higher S/N}. All targets are color-coded based on their adjusted isolation measure, or aperture size, with brighter points being more isolated; targets where signal was detected are in color, whereas non-variable targets are in grayscale. As in Figure~\ref{fig:classifiedcmds}, we do not plot the S/N ratio for non-variable stars.
    \label{fig:ncent}
\end{figure*}

We further compared the noise and S/N noise measurements for the EC-IRIS and TC-IRIS pipeline variants, as outlined in Section~\ref{sec:imscentroids}. In Figure~\ref{fig:ncent}, we show the measured noise and signal-to-noise ratios, with EC-IRIS on the $x$~axis and the ratio of TC-IRIS to EC-IRIS on the $y$~axis. Points are color-coded by isolation measure, showing that noise and S/N per method have no dependence on target isolation. For both noise and S/N, points that lie above the one-to-one line indicate better performance by EC-IRIS. We found that EC-IRIS performed better for both metrics, producing lower noise levels for the majority of targets, and generally higher S/N ratios. The exceptions in S/N tend to be for lower S/N (and therefore less certain) signal detections, where one might expect to find more natural variance in method effectiveness. This leads us to suggest the use of an ensemble centroid measurement as favorable for future applications of image subtraction photometry; it has the additional advantage of saving significant computational time.

\begin{figure*}
    \centering
    \includegraphics[width=\textwidth]{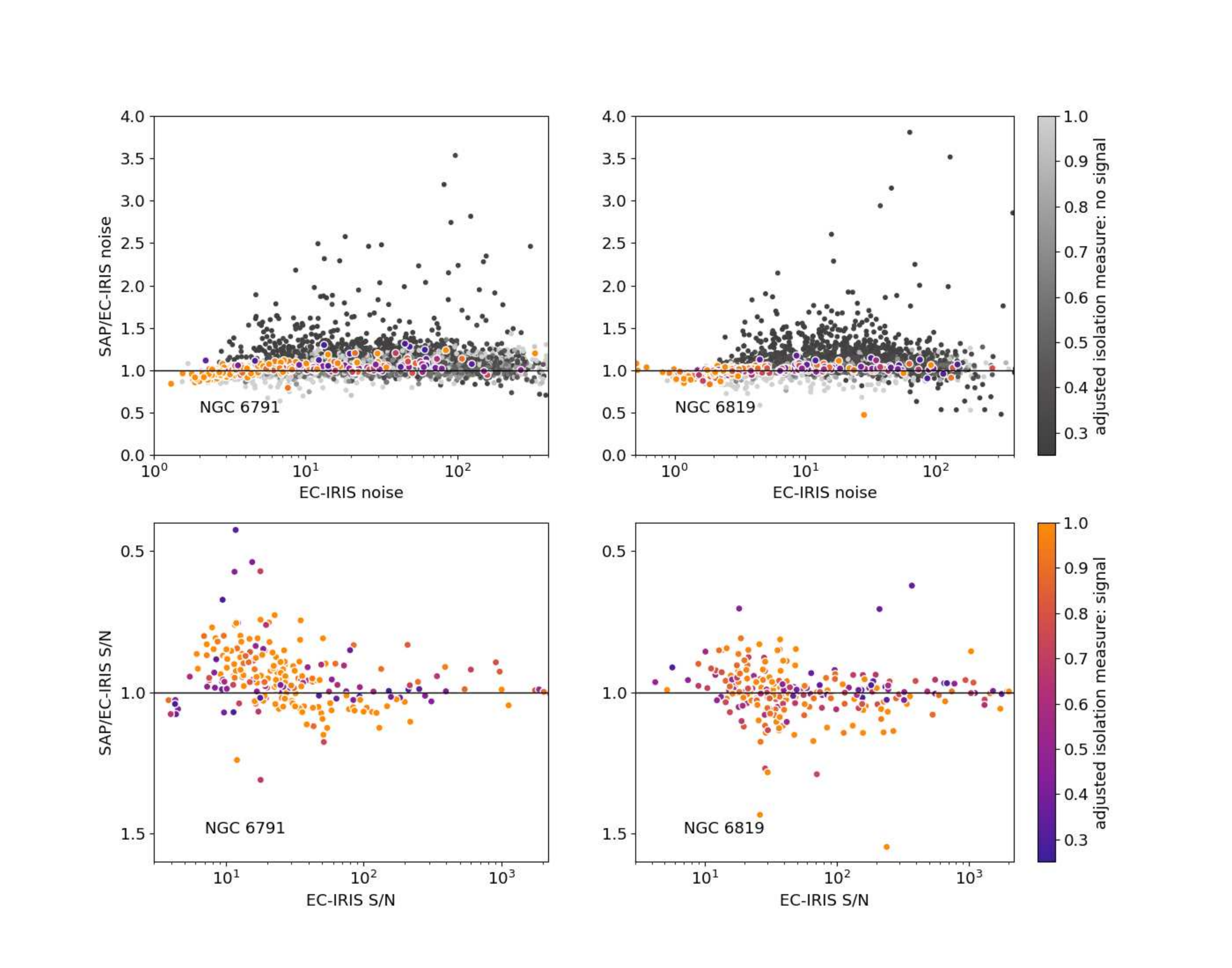}
    \caption{Left panels: NGC~6791. Right panels: NGC~6819. Top row: noise measures, with EC-IRIS image subtraction on the $x$~axis and the ratio between EC-IRIS and custom SAP on the $y$~axis. Bottom row: signal-to-noise measures for EC-IRIS and custom SAP. Points are color-coded as in Figure~\ref{fig:ncent}, and as in Figures~\ref{fig:classifiedcmds} and \ref{fig:ncent}, we do not plot the S/N ratio for non-variable stars.}.
    \label{fig:nsap}
\end{figure*}

We also compared EC-IRIS to the output of a custom SAP pipeline, as described in Section~\ref{sec:imsrunning}. We show this in Figure~\ref{fig:nsap}, plotted with the same $y$-axis limits as the TC-IRIS/EC-IRIS ratio in Figure~\ref{fig:ncent} for clear comparison. We found that SAP produced noisier light curves than EC-IRIS for fainter targets, though not as noisy as TC-IRIS, suggesting that further improvements can be made to the application of a target centroid measurement as part of the IRIS algorithm. EC-IRIS tends to be noisier than SAP for brighter targets. Encouragingly, however, we found that EC-IRIS recorded higher S/N ratios for the majority of detected variable targets, in particular low S/N targets, regardless of their noise levels or isolation. This demonstrates that image subtraction performs as expected in emphasizing light curve variability and, for the purposes of studying a large volume of stars in a crowded field, produces more favorable results than SAP.

\subsection{Comparison to \kep\ data}
\label{sec:lckep}

\begin{figure}
    \centering
    \includegraphics[width=0.6\textwidth]{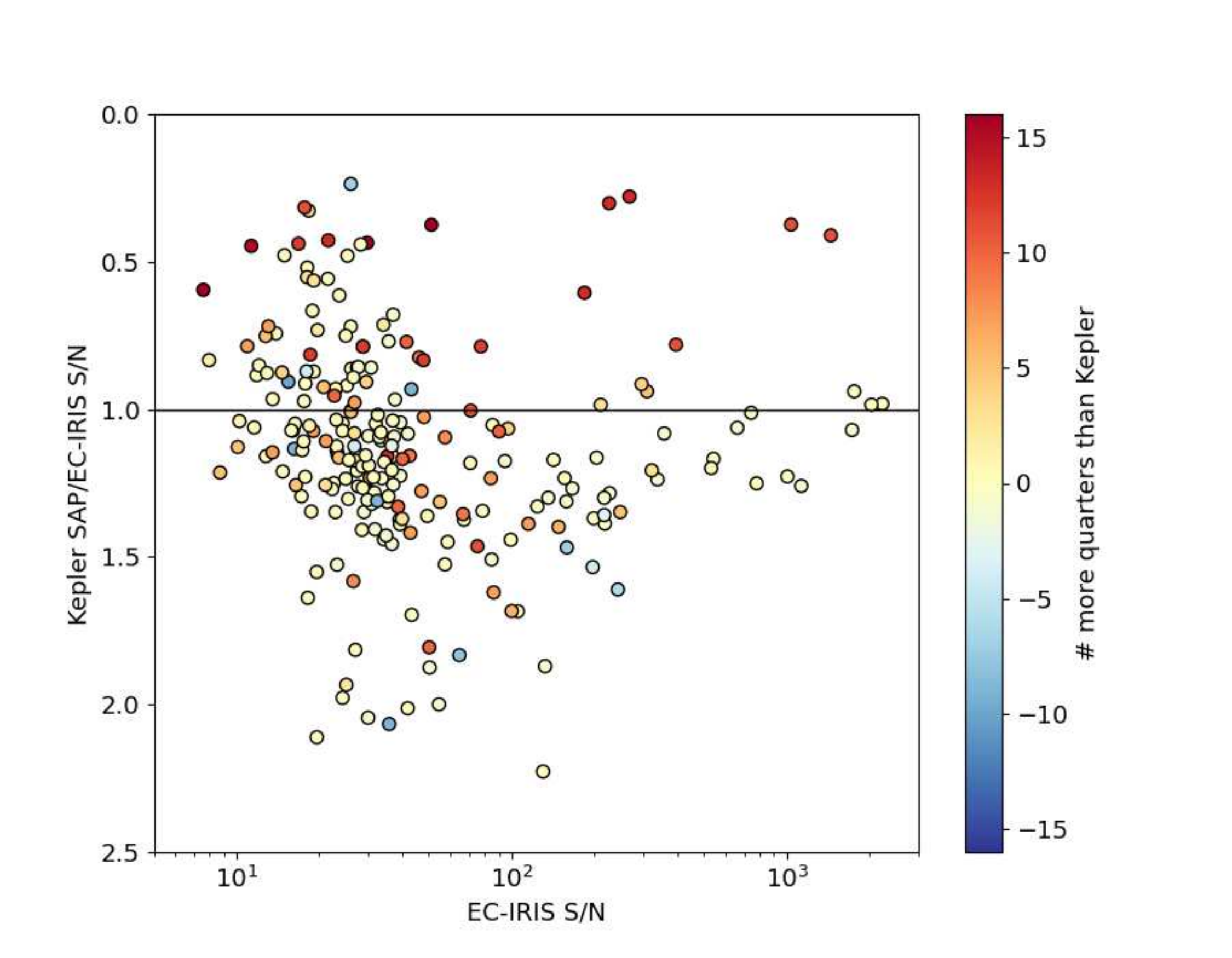}
    \caption{Comparing the retrieved S/N of 230 stars with signal in the sample between EC-IRIS photometry and \kep\ SAP photometry, using the same data preparation methods. Points are colored according to how many more quarters of data are available in the IRIS catalog. The $y$-axis is inverted as in Figures~\ref{fig:ncent} and \ref{fig:nsap} to indicate that points above the line represent a better result from the EC-IRIS pipeline. Note that the $x$-axis is shown in log scale to display outliers with very high S/N measurements.}
    \label{fig:sncomp}
\end{figure}

We compared our data to \kep\ long cadence SAP data for 382 available stars, using the same light curve processing methods to retrieve noise and S/N measures. We detected signal in 230 of these stars, creating a list of targets where S/N is a relevant measure --- we then compared the S/N ratio from EC-IRIS data to the S/N ratio from \kep\ SAP, as shown in Figure~\ref{fig:sncomp}. The data points in this figure are color-coded by how many more quarters of superstamp data are available than of \kep\ long cadence SAP data. For the majority of these stars, we were able to extract more quarters than the stars were targeted for during the \kep\ mission. As would be expected, we find a weak correlation showing that targets for which we have more quarters than \kep\ SAP also have higher S/N ratios in EC-IRIS. However, this is not universally true; for the majority of these stars we retrieve higher S/N ratios from \kep\ SAP.

\begin{figure}
    \centering
    \includegraphics[width=0.8\textwidth]{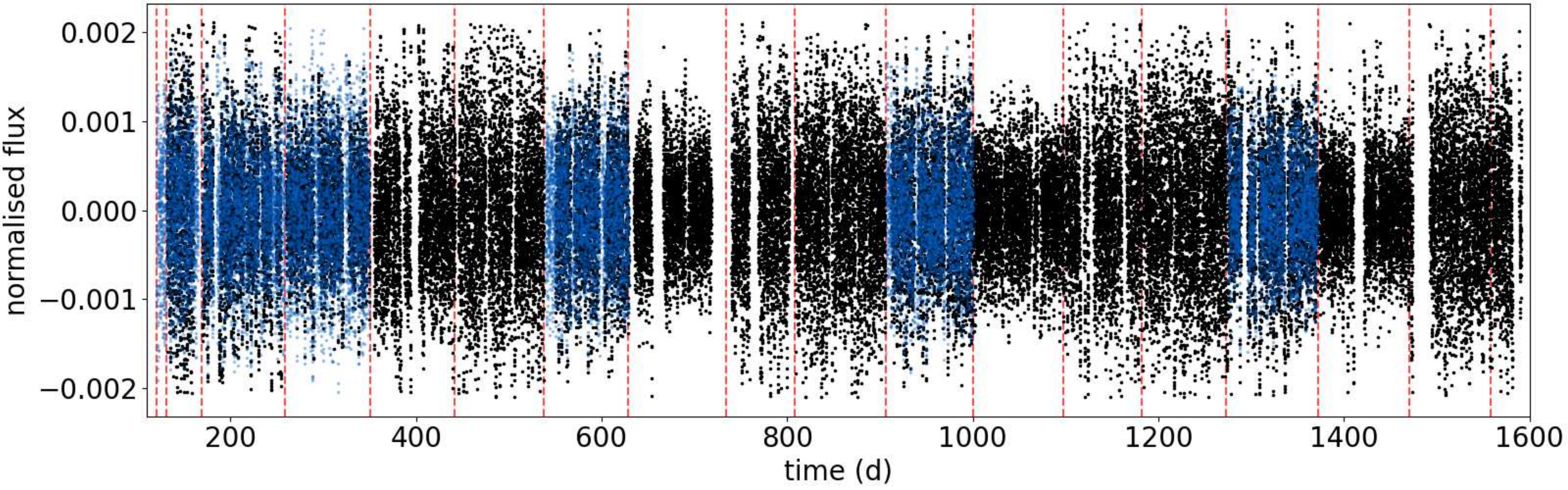}
    \includegraphics[width=0.8\textwidth]{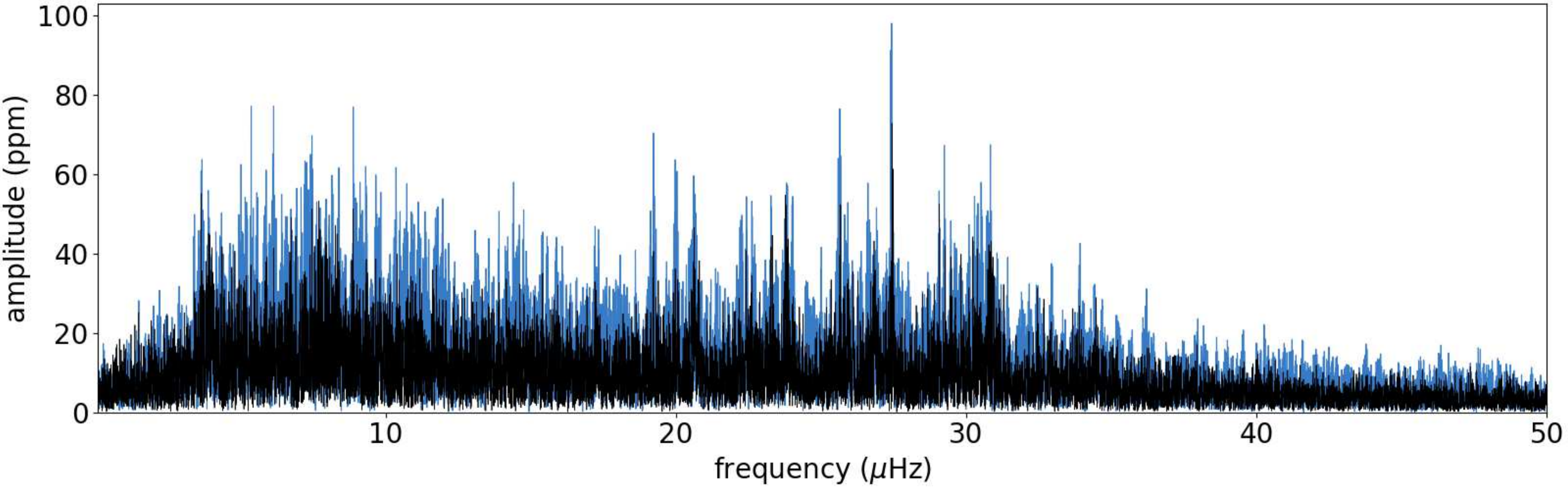}
    \caption{Time series (top) and amplitude spectrum (bottom) for KIC~2570715, an oscillating red giant with only seven available quarters of \kep\ SAP data. EC-IRIS data are shown in black, and \kep\ SAP in blue. Both light curves in the top panel were produced using the same processing methods. Dashed red lines indicate quarter boundaries.}
    \label{fig:2570715}
\end{figure}

There are a number of potential contributing factors for the case where we have more quarters available than \kep\ SAP but the measured S/N ratio from our data is lower. One simple explanation is the differences in photometry and light curve processing methods between the two pipelines. Most of these stars are relatively isolated, i.e. there are no significant crowding issues in the TPFs; the \kep\ SAP pipeline is optimized for these stars, whereas the EC-IRIS pipeline is optimized for faint stars. Additionally, our light curve processing routine is designed to be broadly applicable but not necessarily tailored to individual stars, so some of these targets may be ``bad luck'' cases where the star is less suited to the processing. Finally, we note that our noise measurement is taken between 250--280\muHz, and if the EC-IRIS pipeline noise happened to be higher in that range for any given star, this could lead to a misleadingly low S/N measurement. Nevertheless, for the stars where we increase the number of quarters available compared to \kep\ long cadence SAP, the improvement in quality is significant. Figure~\ref{fig:2570715} shows the time series of KIC~2570715, an oscillating red giant with only 7 quarters of \kep\ long cadence data available. Though the EC-IRIS data (black) suffers from some discontinuity at quarter boundaries, the scatter within each quarter is comparable to the \kep\ SAP (blue) scatter, and amplitude spectra show that the oscillations are much clearer in the EC-IRIS time series.

\section{Discussion and conclusions}
\label{sec:conc}

In this work, we produced light curves for stars in and around the \kep\ open clusters NGC~6791 and 6819, using the \kep\ superstamps, which are 200x200 pixel images with long cadence time series coverage of $\sim$14$''$ square regions around the two clusters. Of the 9,150 light curves included in this data release, 8,427 are new light curves for objects which were not targeted for individual photometry during the \kep\ mission. Using a novel implementation of the image subtraction photometry method, the \kep\ IRIS light curves exhibit low noise levels for faint targets, which comprise the majority of the catalog, and prioritize variable signal. The catalog is available as a high level science product (HLSP) from MAST, and expands the completeness of the \kep\ long cadence dataset, as well as providing new light curves for the further study of two old open clusters.

This study has shown that image subtraction photometry is a highly suitable method for producing light curves for faint targets, and also for performing crowded field photometry. Using a novel algorithm, we show that increased resolution image subtraction (IRIS) is particularly well-suited to working with space-based data. With this initial study of the dataset, we have already made some new finds, including the excess of variable targets in NGC~6819 and a new \& undocumented \kep\ systematic (Section~\ref{sec:lcsys}). We have identified key targets for further study such as KIC~2438249, the \gdor\ eclipsing binary analyzed by \citet{li_effect_2020}.

There are inevitably drawbacks associated with a dataset of this size. We issue two caveats for users of the \kep\ IRIS data. Firstly, the light curves were processed and corrected using a one-size-fits-all method, aimed at maximizing the variability yield without needing to perform custom corrections on individual targets. We corrected using a 100~d Gaussian filter, which necessarily means that longer period signals will be lost, and suggests that we may have slightly underestimated the overall variability fraction across both cluster fields. We strongly recommend that users searching for long-period targets work with the raw light curves. These provide their own challenges, such as differing point scatter between quarters, as outlined in Section~\ref{sec:lcproc}. The second caveat for users is that, due to the crowding of these fields and the emphasis on variability provided by image subtraction photometry, signal blending is inevitable in the light curves of targets adjacent to variable stars which are particularly subject to crowding. We suggest that variability should be compared with targets nearby in RA and DEC in order to confirm the source of the signal.

In future work, we aim to adapt IRIS for \T\ data. The \T\ full frame images cover almost the whole sky, including a wealth of open clusters in the field, providing many opportunities for crowded field photometry. The fields around open clusters would also provide a useful way to test the brightness and isolation limits of the efficacy of this method for \T\ data. For future use, we can improve on the TC-IRIS method, which appears to be underperforming here, given that a comparable SAP procedure produces better results. Improved TC-IRIS will be useful for \T\ data, as it quickly becomes unfeasible to produce ensemble centroid measurements for data of that scope. We can also explore alternative methods, such as performing cross-correlation across the superstamp time series. Another potential use for the IRIS pipeline is with regular long or short cadence \kep\ TPFs, aiming to improve signal-to-noise ratios for faint and/or crowded targets, or to extract light curves for serendipitous background stars in larger TPFs.

Finally, we note that there is a wealth of science that can be done using this data release, such as testing new light curve correction methods to search for long period variability, studying and modeling the large range of variable stars across both cluster fields, and searching for exoplanets. It is our hope that these data will be of broad use to the community as both an expansion of the \kep\ long cadence dataset and a new avenue for studying the open clusters NGC~6791 and NGC~6819.

\section*{Acknowledgements}

ILC acknowledges scholarship support from the University of Sydney and the Hunstead Student Support Scholarship. Much of this paper is drawn from, and elaborates on, Chapters 4 and 5 of ILC's PhD thesis \citep{colman_thesis_2020}. ILC acknowledges Jason Drury for providing the cluster memberships that were used for target selection in the initial study; Gang Li for assisting with the analysis of several \gdor\ targets featured in her thesis; thesis referees Aliz Derekas, Frank Grundahl, and Susan Mullally for their feedback and comments on this work; and S. Mullally, Stephen Bryson, Douglas Caldwell, and Karsten Brogaard for further discussion and input on this project. We gratefully acknowledge support from the Australian Research Council (DP210103119) and from the Danish National Research Foundation (Grant DNRF106) through its funding for the Stellar Astrophysics Center (SAC).


\software{Astropy \citep{collaboration_astropy:_2013, collaboration_astropy_2018},
Pandas \citep{reback_pandas_2021},
NumPy \citep{harris_array_2020},
SciPy \citep{virtanen_scipy_2020},
Lightkurve \citep{collaboration_lightkurve:_2021},
Matplotlib (http://dx.doi.org/10.1109/MCSE.2007.55),
GNU Parallel \citep{tange_ole_2018_1146014}
}

\bibliography{sample631}{}
\bibliographystyle{aasjournal}



\end{document}